\documentclass[sigconf]{acmart}

\usepackage{balance}
\usepackage{fancyhdr}
\usepackage{graphicx}
\graphicspath{{images/}}

\usepackage{subfigure}
\usepackage{url}
\usepackage{epstopdf}
\usepackage{algorithm}
\usepackage{algorithmic}
\usepackage{makecell}
\newcommand{\tabincell}[2]{\begin{tabular}{@{}#1@{}}#2\end{tabular}}
\def\bR{\mathbb{R}}

\def\bZ{\mathbb{Z}}

\def\cD{\mathcal{D}}

\def \qed {\hfill \vrule height6pt width 6pt depth 0pt}
\def\bee{\begin{equation}}
\def\ene{\end{equation}}
\def\been{\begin{equation*}}
\def\enen{\end{equation*}}

\newtheorem{pro}{Proposition}[section]
\newtheorem{lem}{Lemma}[section]

\renewcommand\footnotetextcopyrightpermission[1]{} 
\AtBeginDocument{%
  \providecommand\BibTeX{{%
    \normalfont B\kern-0.5em{\scshape i\kern-0.25em b}\kern-0.8em\TeX}}}

\setcopyright{acmcopyright}
\copyrightyear{2018}
\acmYear{2018}
\acmDOI{10.1145/1122445.1122456}

\acmConference[Woodstock '18]{Woodstock '18: ACM Symposium on Neural
  Gaze Detection}{June 03--05, 2018}{Woodstock, NY}
\acmBooktitle{Woodstock '18: ACM Symposium on Neural Gaze Detection,
  June 03--05, 2018, Woodstock, NY}
\acmPrice{15.00}
\acmISBN{978-1-4503-XXXX-X/18/06}

\begin{document}

\title{Dynamic Pricing for Client Recruitment in Federated Learning}





\author{Xuehe~Wang}
\affiliation{
    \institution{School of Artificial Intelligence \\ Sun Yat-sen University,Zhuhai,}
    \postcode{519082}
    \institution{and the Guangdong Key Laboratory of Big Data Analysis and Processing}
    \city{Guangzhou}
    \postcode{510006}
    \country{China}
}
\email{wangxuehe@mail.sysu.edu.cn}

\author{Shensheng~Zheng}
\affiliation{
    \institution{School of Artificial Intelligence \\ Sun Yat-sen University}
    \city{Zhuhai}
    \postcode{519082}
    \country{China}
}
\email{zhengshsh7@mail2.sysu.edu.cn}

\author{Lingjie~Duan}
\affiliation{
    \institution{Pillar of Engineering Systems and Design,  Singapore University of Technology and Design}
    \postcode{487372}
    \country{Singapore}
}
\email{lingjie\_duan@sutd.edu.sg}

\renewcommand{\shortauthors}{Xuehe et al.}

\begin{abstract}
Though federated learning (FL) well preserves clients' data privacy, many clients are still reluctant to join FL given the communication cost and energy consumption in their mobile devices.  It is important to design pricing compensations to motivate enough clients to join FL and distributively train the global model. Prior pricing mechanisms for FL are static and cannot adapt to clients' random arrival pattern over time. We propose a new dynamic pricing solution in closed-form by constructing the Hamiltonian function to optimally balance the client recruitment time and the model training time, without knowing clients' actual arrivals or training costs. During the client recruitment phase, we offer time-dependent monetary rewards per client arrival to trade off between the total payment and the FL model's accuracy loss. Such reward gradually increases when we approach to the recruitment deadline or have greater data aging, and we also extend the deadline if the clients' training time per iteration becomes shorter. Further, we extend to consider heterogeneous client types in training data size and training time per iteration. We successfully extend our dynamic pricing solution and develop an optimal algorithm of linear complexity to monotonically select client types for FL. Finally, we also show robustness of our solution against estimation error of clients' data sizes, and run numerical experiments to validate our conclusion.

\end{abstract}




\maketitle

\section{Introduction}

The Internet of Things (IoT) is fast developing and well connects many human clients when using their mobile devices. It is desirable to learn from the massive data generated on such mobile devices to train the client models (e.g., for personal advertisement or recommendation) \cite{NEURIPS2020_e32cc80b}. To preserve clients' data privacy, federated learning (FL) is proposed to invite clients to iteratively update their computing results using local data, without sharing such data to the cental server \cite{yang2019federated}. In a FL system (e.g., Gboard \cite{hard2018federated}),  a learning task is implemented in two phases: client recruitment and model training. In the client recruitment phase, the central server waits for sequential arrivals of mobile clients to connect and participate; in the model training phase, clients obtain the central server's aggregation parameter feedback in last iteration and compute their local results for this iteration to update. In this phase, the central server well utilizes clients’ computing power and local data to train the desired model \cite{tran2019federated}.



Previous works mainly focus on the technological issues of FL. For example, to improve communication efficiency for FL, there are local updating \cite{mcmahan2017communication,smith2018cocoa}, compression schemes \cite{konevcny2016federated, caldas2018expanding} and decentralized training \cite{reisizadeh2019robust, he2018cola}. To enhance the overall security and privacy of FL systems, Fung et al. \cite{fung2018mitigating} propose FoolsGold that identifies poisoning sybils based on the diversity of client contributions in the distributed learning process. Liu et al. \cite{liu2020secure} further introduce a blockchain-based secure FL framework to prevent malicious or unreliable participants in FL. Hao et al. \cite{hao2019towards} propose an efficient and privacy-preserving federated deep learning protocol based on stochastic gradient descent method, by integrating the additively homomorphic encryption with differential privacy.

Most of their works assume that the clients will voluntarily participate in FL, which may not be realistic due to clients' training cost including computational energy consumption on model training and parameter transmission.
In reality, many clients are reluctant to participate in model training if there is not enough compensation during the training process. Thus, to increase clients engagement and ensure training accuracy, it is important to design pricing compensations to motivate enough clients to join FL and distributively train the global model. There are only a few recent works discussing the incentive mechanism design in FL. Ding et al. \cite{ding2020incentive} use contract-theoretic approach to best trade off between client contribution for FL and total payment, by considering the clients' multi-dimensional private information. Some other researchers (e.g., Feng et al. \cite{feng2019joint}, Kang et al. \cite{kang2019incentive} and Sarikaya et al. \cite{sarikaya2019motivating}) formulate a Stackelberg game to design pricing incentive against the clients' following responses. Zhan et al. \cite{zhan2020big} design an incentive mechanism to optimize the utilities of mobile clients and accuracy of the training model by considering the different sensing and training capabilities of mobile clients. Reinforcement learning is also used to derive the optimal pricing strategy for the central server to recruit clients for training \cite{zhan2020learning, zhan2020incentive}.

However, there are several overlooked points in the above mechanisms to fit the FL scenario.  Firstly, the above literature simply decides static pricing strategies for client recruitment, by assuming all potential clients are always waiting there for FL tasks. In many FL practices (e.g., Gboard \cite{hard2018federated}), mobile clients have their own timing or face message delay, randomly arriving over time to participate. In dynamic case, one-shot or static pricing can easily lead to data over sampling or under sampling, and it is important to adaptively adjust the pricing compensation based on clients' actual arrival pattern and cost distribution to meet the data target for later model training. Secondly, the clients' training costs or even their arrival pattern are unknown to the FL system, and clients in general have different data sizes to contribute and training time per iteration in model training. Most of the incentive mechanisms (e.g., \cite{feng2019joint, kang2019incentive, sarikaya2019motivating}) assume that the FL system has complete information about the clients without any uncertainty to operate. In the few works considering the incomplete information about clients' private training costs and random arrivals, they mainly use static strategy to model the incentive mechanism \cite{8716560, ding2020incentive}. Therefore, in the client recruitment phase, it is necessary to design a dynamic pricing strategy to incentivize heterogeneous clients to participate in FL under incomplete information. Moreover, for finite time horizon, we need to balance the client recruitment time and the model training time per task, while the incentive works above mainly focus on the first phase of client recruitment. Finally, when facing multiple types of clients with different data size and training time per iteration, the client-type selection should be further taken into consideration to trade off the total data size and global training iterations. 

Our key novelty and main contributions are summarized as follows:
\begin{itemize}
\item \emph{Dynamic client recruitment in federated learning:}
To our best knowledge, this paper is one of the first works studying how to motivate dynamically arriving clients to participate in the federated learning over time, without knowing their private training costs or even arrival pattern. Due to the non-trivial waiting time for enough clients, we jointly consider two phases to balance for each FL task: client recruitment and the model training by the involved clients, where the longer time for recruitment helps gain more training data yet leaves less time for FL convergence.

\item \emph{Dynamic pricing mechanism under incomplete information:}
In the client recruitment phase, we decide time-dependent monetary rewards in closed-form by constructing the Hamiltonian function to balance the total payment to clients and model accuracy loss, where a higher pricing reward attracts more clients for data contribution to FL yet adds expense to the system. We prove that the central server should provide a higher price when approaching to the recruitment deadline or with greater data aging, and show our dynamic pricing strategy always outperforms the static pricing strategy.



\item \emph{Threshold-based client recruitment policy to cope with the dynamic pricing scheme:}
Though a longer client recruitment duration helps recruit more clients and enlarge the training dataset, it leaves less model training time. After deciding the dynamic pricing for any recruitment time, we systematically analyze the best partition in client recruitment time and model training time, which is proved to be threshold-based. As compared to model training, we relatively reserve less time for client recruitment given greater data aging or longer training time of each client.

\item \emph{Pricing extension to heterogeneous clients with robustness check:}
We extend the pricing solution to consider heterogeneous client types in training data size and training time per iteration. Though a client type may have a greater data size, it also incurs greater computing time to accommodate in the synchronous training. We successfully extend our dynamic pricing solution and develop an optimal algorithm of linear complexity (with respect to the number of client types) to monotonically select client types for FL. Note that given a selected client type, the iteration duration is fixed and including client types with smaller data size and training time only creates updates within the iteration duration without reducing the number of global training iterations. Thus we prove that the optimal client-types should be selected monotonically to accelerate the model training. We also show robustness of our solution even if we have some error in estimating clients' data size. We also run experiments to validate our conclusion.
\end{itemize}

The rest of this paper is organized as follows. The system model and problem formulation are given in Section \ref{sec_systemmodel}. In Sections \ref{sec_optPricing} and \ref{sec_optThreshold}, we analyze the optimal dynamic pricing and data recruitment threshold for homogeneous clients. The extension to heterogeneous clients case and the robustness checking are discussed in Section \ref{sec_mutiType}. Experimental results are shown in Section \ref{sec_simulations}. Section \ref{sec_conclusion} concludes this paper.

\section{System Model and Problem Formulation}\label{sec_systemmodel}

\begin{figure}[ht]
\centering\includegraphics[scale=0.30]{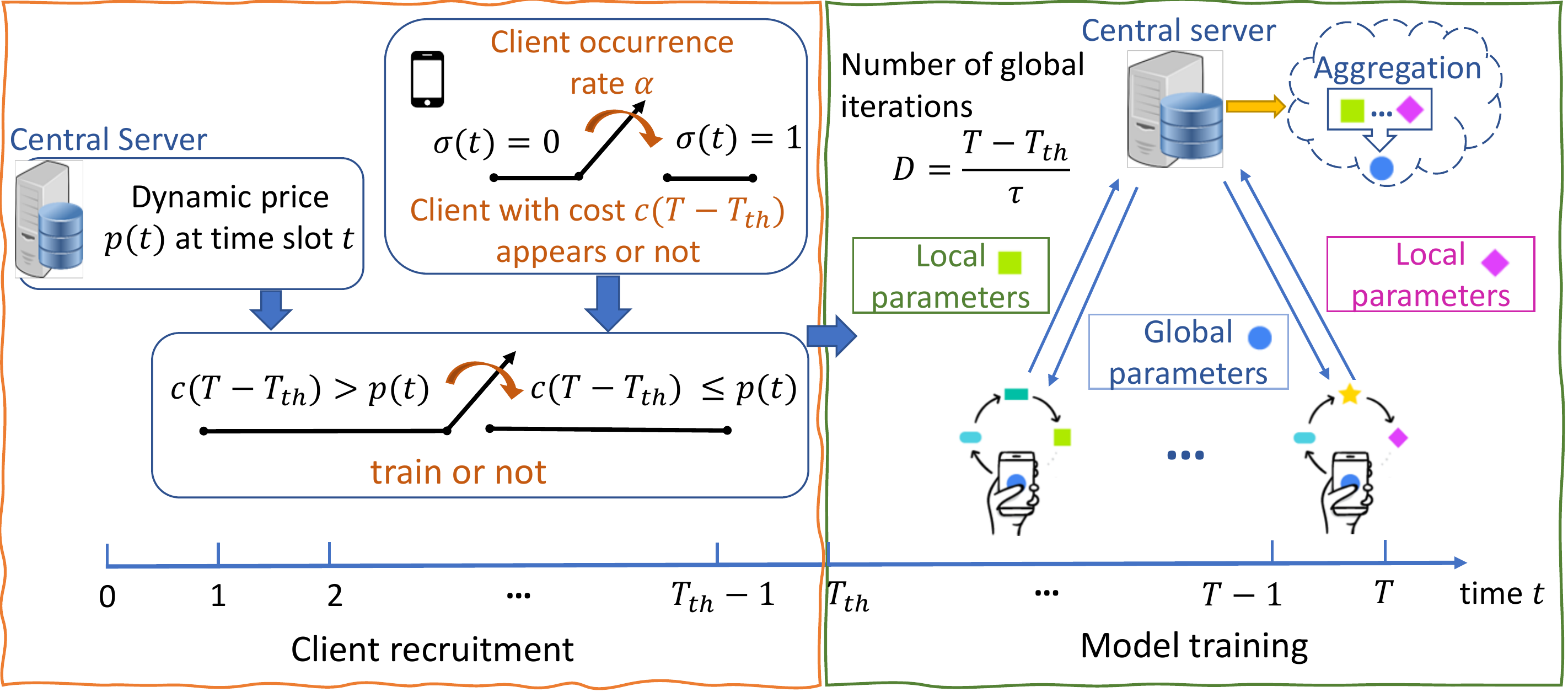}
\caption{Illustration of the two-phase model including the client recruitment phase and model training phase in federated learning.}
\label{fig_systemmodel}
\end{figure}


We consider a FL platform that plans to complete a task in $T$ time slots over the discrete time horizon. It first recruits dynamically arriving clients in the first $T_{th}$ time slots and then proceeds with model training at recruited clients' mobile devices using their local data in the rest $T-T_{th}$ slots, as shown in Fig. \ref{fig_systemmodel}. Note that in client recruitment phase, the agreed clients just commit to help for the later model training, without incurring any cost. They only incur costs in computational energy consumption and parameter transmission during the model training phase. We first discuss homogeneous clients with identical data size and training time. The extension to heterogeneous clients with different data size and training time will be presented in Section \ref{sec_mutiType}. For ease of reading, we list the key notations in Table \ref{tab:Key_Notations}.

\begin{table}[ht!]
  \begin{center}
    \caption{Key notations and their physical meanings.}
    \label{tab:Key_Notations}
    \begin{tabular}{|l|l|} 
      \hline
      $T$ &  Total time horizon \\
      \hline
      $T_{th}$ & Time threshold for client recruitment phase \\
      \hline
      $D$ &  Number of global iterations \\
      \hline
      $N$ &  Number of heterogeneous client types \\
      \hline
      $c$ &  Client's unit cost per training time $\in [0, b]$ \\
      \hline
      $\alpha$ & Arrival rate of the clients in each time slot \\
      \hline
      $r$ &  Discount factor to indicate the freshness of data \\
      \hline
      $s$ &  Data size of the homogeneous clients \\
      \hline
      $s_{i}$ & Data size of the $i$-th client type \\
      \hline
      $\tau$ & \tabincell{l}{Training time per iteration for \\homogeneous clients} \\
      \hline
      $\tau_{i}$ & Training time per iteration of the $i$-th client type \\
      \hline
      $p(t)$ &  \tabincell{l}{Recruitment price of the homogeneous clients\\ at time slot $t \in \{0,,,,,T_{th} - 1\}$} \\
      \hline
      $p_{i}(t)$ &  \tabincell{l}{Recruitment price of the $i$-th client type \\ at time slot  $t \in \{0,,,,,T_{th} - 1\}$} \\
      \hline
      $q_i$  & Percentage of type-$i$ clients with $\sum_{i=1}^Nq_i=1$ \\
      \hline
      $B(t)$ & \tabincell{l}{Resulting total data size at time slot \\ $t \in \{0,,,,,T_{th}\}$ } \\
      \hline
      $U(T)$ &  Total expected cost of the homogeneous clients \\
      \hline
      \tabincell{l}{$J(\boldsymbol{P(t)}|T_{th},$\\ $\{1,...,j\})$} &  \tabincell{l}{Total expected cost given client-types \\ $\{1, ..., j\}$ and recruitment threshold $T_{th}$} \\
      \hline
    \end{tabular}
  \end{center}
\end{table}

\subsection{Client Recruitment Phase}\label{data_rec_phase}
For the client recruitment phase, the central server recruits data from clients for the FL task in any time slot $t=0,...,T_{th}-1$. As shown in Fig \ref{fig_systemmodel}, at the beginning of each time slot $t\in\{0,...,T_{th}-1\}$, the central server announces price $p(t)$ for training $T-T_{th}$ time slots. Then a client may appear randomly in this time slot and (if so) he further decides to help train the model or not by comparing the price offer $p(t)$ and his own total cost $c(T-T_{th})$, where $c$ is the unit cost per training time \cite{AWS}.
Considering the clients appear according to a Poisson process,  with an average number $\lambda$ of client arrivals per unit time. Then, the probability of the random number of client arrivals $N(t)$ in the $t$-th time slot of the interval $[(t-1)\Delta, t\Delta)$ being equal to $k$ is
\bee Pr(N(t)=k)=\frac{e^{-\lambda\Delta}(\lambda\Delta)^k}{k!}, \ene
with time duration $\Delta$ for each time slot. Note that $Pr(N(t)>1)$ becomes trivial as long as the time duration $\Delta$ is small. Thus, each time slot's duration $\Delta$ is properly selected to be short such that it is almost sure to have at most one client arrival at a time.
Then, whether a client appears at time slot $t$ is represented as $\sigma(t)$:
\begin{equation}
\sigma(t) =
\begin{cases}
1,& \text{if a client arrives in time slot } t,\\
0,& \text{otherwise,}
\end{cases}
\end{equation}
with the client arrival rate in each time slot $\alpha=\lambda\Delta$.

We consider that the clients' private costs are i.i.d. according to a cumulative distribution function (CDF) $F(c), c\in[0,b]$, where upperbound $b\geq 1$ is estimated from historical data. Though all potential users' costs follow the same distribution, their realized costs are different in general. 
If a client with data size $s$ and training time $\tau$ per iteration appears and accepts the price $p(t)$ at time slot $t$, i.e., $\sigma(t) = 1$ and $c(T-T_{th}) \leq p(t)$, he will help train the model using his local training dataset and return the updated local parameter within the training time $\tau$, where $\tau$ represents the clients' training time (total computation and transmission time) per global iteration. The payoff of the participating client with private unit cost $c$ is the difference between the price and his own training cost, i.e., $p(t) - c(T-T_{th})$. Then, the arrival client's payoff at time $t$ is concluded as follows:
\begin{equation}\label{equ_clientutility}
\Upsilon(t) =
\begin{cases}
p(t) - c(T-T_{th}), &  \sigma(t) = 1 \text{ \& } c\leq\frac{p(t)}{T - T_{th}}, \\
 0, & \text{otherwise. }
\end{cases}
\end{equation}

In this paper, we consider non-trivial client arrival rate $\alpha\geq 0.5$ and training efficiency (data size/training time) $\frac{s}{\tau}\geq 1$. Otherwise, the central server may always set trivial price $p(t)$ to be the upperbound $b(T-T_{th})$ to include all arriving clients without missing critical data.

\subsection{Model Training Phase}\label{model_train_phase}

For distributed model training phase, we consider the synchronous FL with the one-step local update, which means each client performs one step of mini-batch stochastic gradient decent (SGD) to update the model parameters in each round, and the server waits for all clients' local parameter updates and then sends the updated global parameter to all clients at the same time for next round's training. Given $M$ clients agreed to participate, each participating client $n \in \{1, ..., M\} $ uses its local dataset $\cD_n$ with data size $\varsigma_n$ to train the model. Denote the collection of data samples in $\cD_n$ as $\{x_k,y_k\}_{k=1}^{\varsigma_n}$, where $x_k\in\bR^d$ is the input sample vector and $y_k\in\bR$ is the labeled output value for the sample $x_k$ at client $n$.
For a sample data $\{x_k,y_k\}$, the objective is to find the model parameter $\omega\in\bR^d$ that predicts the output $y_k$ based on $x_k$ with the loss function $f_k(\omega)$, where $f_k(\omega)$ characterizes the difference between the predicted value $\hat{y}_k(x_k,\omega)$ and real output $y_k$. The loss function on the data set $\cD_n$ of client $n$ is
\begin{equation*}
F_n(\omega)=\frac{1}{\varsigma_n}\sum_{k\in\cD_n}f_k(\omega).
\end{equation*}
At each iteration $t+1$, client $n$ updates its local parameter based on last global parameter $\omega_t$ sent by the central server:
\begin{equation*}
\omega_{t+1}^n=\omega_{t}-\eta\nabla F_n(\omega_{t}),
\end{equation*}
where $\eta$ is the learning rate, and sends $\omega_{t+1}^n$ back to the central server. The server averages the parameters sent back by $M$ participating clients
\begin{equation*}
\omega_{t+1}=\sum_{n=1}^M\frac{\varsigma_n}{\varsigma}\omega_{t+1}^n,
\end{equation*}
where $\varsigma=\sum_{n=1}^M\varsigma_n$ is the total data size, and sends the updated global parameter $\omega_{t+1}$ to all clients for next round's training.


The optimal model parameter $\omega^*$ that minimizes the global loss function is:
\begin{equation*}
\omega^*=\arg\min_{\omega}f(\omega)=\arg\min_{\omega}\sum_{n=1}^M\frac{\varsigma_n}{\varsigma}F_n(\omega).
\end{equation*}

Note that there are $T-T_{th}$ time slots left for training and the number of training iterations is $D=\frac{T - T_{th}}{\tau}$. The accuracy loss after $D$ global iterations is measured by the difference between the global loss with the predicted parameter $\omega_D$ and that with the optimal parameter $\omega^*$, i.e., $f(\omega_D)-f(\omega^*)$. The expected model accuracy loss is $O(\frac{1}{\sqrt{B(T_{th})D}}+\frac{1}{D})$ \cite{Dekel2012Optimal}, where $B(T_{th})$ is the total data size contributed by the participating clients at the end of the client recruitment phase. For finite time horizon $T$, the objective $U(T)$ of the central server is to find the optimal dynamic pricing $p(t), t \in\{0,\dots,T_{th}-1\}$ and recruitment threshold $T_{th}$ to minimize the total expected cost, which is the summation of the total expected payment to clients and the expected model accuracy loss:

\bee \label{equ_U(T)1}
U(T) = \min_{
\substack{p(t), t \in\{0,\dots,T_{th}-1\}\\ T_{th}\in\{1,...,T-1\}}}
\sum_{t=0}^{T_{th}-1}\zeta(p(t))+\frac{1}{\sqrt{B(T_{th})D}}+\frac{1}{D}, \ene
where $\zeta(p(t))$ is the expected payment at time $t$.

\subsection{Problem Formulation}
Based on the discussions in Sections \ref{data_rec_phase} and \ref{model_train_phase}, we use dynamic Bayesian Game to model our federated learning system's interaction with dynamically arriving clients without knowing their arrivals and private training costs.
\begin{itemize}
  \item Players: the central server and random arriving clients.
  \item Strategies: For the central server, it decides the dynamic pricing $p(t), t \in\{0,\dots,T_{th}-1\}$, recruitment threshold $T_{th}$ (for heterogeneous clients case, it further decides the client-type choice). For the clients, they decides to help train the model or not.
  \item Information set: The central server does not know clients' arrivals and private costs, but knows the client arrival rate $\alpha$ in each time slot and the cost distribution $F(c)$.
  \item Expected payoffs: The client's payoff $\Upsilon(t)$ in (\ref{equ_clientutility}) depends on his private training cost $c$ and price $p(t)$. The central server's total cost $U(T)$ in (\ref{equ_U(T)1}) is the summation of the total payment to clients and model accuracy loss.
\end{itemize}


Note that a longer recruitment duration may help recruit more clients, but leaves less training time under finite time horizon $T$. Thus, the central server should balance the client recruitment time and model training time. Moreover, in the client recruitment phase, static pricing cannot adapt to clients’ random arrival patterns and data aging over time. Hence, we need to use dynamic pricing to balance the total payment to clients and the accuracy loss under incomplete client information.
By considering the above issues, we formulate the problem by two-stage:
\begin{itemize}
  \item Stage I: The central server chooses the optimal client recruitment threshold $T_{th}$.
  \item Stage II: Given the optimal recruitment threshold $T_{th}$, the central server decides the optimal dynamic pricing $p(t),t\in\{0,\dots,T_{th}-1\}$.
\end{itemize}

In the following, we use backward induction to first analyze the dynamic pricing in Stage II given the time threshold, then the optimal recruitment threshold in Stage I.


%

\section{Optimal Dynamic Pricing under Incomplete Information in Stage II}\label{sec_optPricing}

In this section, we study the central server's pricing strategy under incomplete information, i.e., the central server does not know the clients' arrivals during the client recruitment phase and the participating client's particular cost. In reality, when waiting for the central server's client recruitment, the value of the data may decrease, e.g., the data collected earlier is not as fresh as the latest data contributed by recently arrived clients. Thus, the useful data size from each device reduces. Here we reasonably consider a non-trivial $0.5 < r < 1$ to model data aging, as a trivial $r$ will lead to no recruitment till the deadline, and our dynamic pricing reduces to one-shot pricing.



For the homogeneous clients with identical data size $s$ and training time $\tau$, from the initial training data size $B(t=0)=0$, the probability that the data size $B(t+1)$ at time $t+1$ increases to $B(t)+s$ is $\alpha F(\frac{p(t)}{T - T_{th}})$, i.e., a client appears and accepts the price offer $p(t)$ at time $t$. The probability that the data size $B(t+1)$ at time $t+1$ remains $B(t)$ is $1-\alpha F(\frac{p(t)}{T - T_{th}})$. Consider uniform distribution of the clients' private costs \footnote{Though more involved, our analysis and key results can be extended to some other distributions such as normal.}, i.e., $F(c)=\frac{c}{b},c\in[0,b]$, the dynamics of the expected data size $B(t)$ is given as:
\bee\label{equ_BUniform}\begin{split} B(t+1)=&r((B(t)+s)\alpha F(\frac{p(t)}{T - T_{th}})\\&+B(t)(1-\alpha F(\frac{p(t)}{T - T_{th}}))\\
=&r(B(t)+\frac{\alpha s}{b(T - T_{th})} p(t)). \end{split}\ene

Since the probability that a client appears and accepts the price $p(t)$ at time $t$ is $\frac{\alpha}{b(T - T_{th})}p(t)$, the expected payment to this client is $\frac{\alpha}{b(T - T_{th})}p^2(t)$. Note that the optimal price $p(t)$ should not exceed the maximum cost $b(T-T_{th})$ of the client as it is unnecessary for the provider to over-pay. Therefore, given any time threshold $T_{th}$, the objective function (\ref{equ_U(T)1}) of the central server can be rewritten as:

\begin{equation}\label{equ_U(T)}\begin{split}
U(T) = \min_{
\substack{p(t)\leq b(T - T_{th})\\ t \in\{0,\dots,T_{th}-1\}}}
&\sum_{t=0}^{T_{th}-1}\frac{\alpha}{b(T - T_{th})}p^2(t)+\frac{1}{\sqrt{B(T_{th})D}}\\&+\frac{1}{D},
\end{split}\end{equation}
\rightline{s.t.\quad\quad$B(t+1)=r(B(t)+\frac{\alpha s}{b(T - T_{th})} p(t))$\quad\quad\quad\quad\quad\quad (\ref{equ_BUniform})}



We can see that a higher price in the client recruitment phase leads to smaller accuracy loss for FL, but cause higher payment to afford for the central server. It's not easy to solve the above problem by considering the huge number of price combinations over time, with computation complexity $O((b(T-T_{th})/\epsilon)^{T_{th}})$ increasing exponentially in $T_{th}$, where $\epsilon$ is the precision of searching for pricing in the range $[0,b]$. In the following proposition, we solve the dynamic pricing in closed-form by constructing the Hamiltonian function. 

\begin{pro}\label{pro_dynamicpricing} The optimal closed-form dynamic pricing $p(t), t\in\{0,\dots$, $T_{th}-1\}$ is given by
\bee\label{equ_p(t)withoutB(t)} p(t)=
\Big(\frac{b^3\tau^3D^2r^{5T_{th}-5t-6}(1-r^2)^3}{16\alpha^3s(1-r^{2T_{th}})^3}\Big)^{\frac{1}{5}}, \ene
which is monotonically increasing in $t$, $p(t)\leq b(T-T_{th})$ holds for any $t\in\{0,\dots,T_{th}-1\}$. The recruited clients' expected data size at the end of the client recruitment phase $T_{th}$ is
\bee\label{equ_BTth} B(T_{th})=D^{-\frac{3}{5}}\big(\frac{\alpha s^2}{4b\tau}r^{2}\sum_{i=1}^{T_{th}}r^{2(i-1)}\big)^{\frac{2}{5}}. \ene

\end{pro}



\textbf{Proof:} According to the problem (\ref{equ_BUniform})-(\ref{equ_U(T)}), we have the discrete time Hamiltonian function as
\bee\begin{split} H(t)=&\frac{\alpha}{b(T-T_{th})}p^2(t)+\lambda(t+1)((r-1)B(t)\\&+\frac{r\alpha s}{b(T-T_{th})} p(t)).\end{split}\ene

Since $\frac{\partial^2 H(t)}{\partial p^2(t)}>0$, the Hamiltonian function $H(t)$ is convex in $p(t)$. Therefore, in order to find the optimal dynamic pricing that minimize the total expected cost $U(T)$ in (\ref{equ_U(T)}), it is necessary to satisfy:
\bee\label{equHwithp} \frac{\partial H(t)}{\partial p(t)}=0, \ene
\bee\label{equ1} \lambda(t+1)-\lambda(t)=-\frac{\partial H(t)}{\partial B(t)}, \ene
with boundary condition
\bee \lambda(T_{th})=\frac{\partial (\frac{1}{\sqrt{B(T_{th})D}}+\frac{1}{D})}{\partial B(T_{th})}=-\frac{1}{2}D^{-\frac{1}{2}}(B(T_{th}))^{-\frac{3}{2}}. \ene

According to (\ref{equ1}), we have $\lambda(t)=r\lambda(t+1)$. Then, based on the boundary condition, we can derive
\bee\label{equ_lambda(t)} \lambda(t)=-\frac{1}{2}r^{T_{th}-t}D^{-\frac{1}{2}}(B(T_{th}))^{-\frac{3}{2}}. \ene
Based on (\ref{equHwithp}) and (\ref{equ_lambda(t)}),
\bee\label{equ_pt_lambda} p(t)=-\frac{rs}{2}\lambda(t+1)=\frac{s}{4}r^{T_{th}-t}D^{-\frac{1}{2}}(B(T_{th}))^{-\frac{3}{2}}. \ene

\noindent Insert $p(t)$ in (\ref{equ_pt_lambda}) into (\ref{equ_BUniform}), for $t\in\{1,\dots,T_{th}\}$, we have
\bee\label{equ_BwithB(T)} B(t)=\frac{\alpha s^2}{4b\tau}D^{-\frac{3}{2}}(B(T_{th}))^{-\frac{3}{2}}r^{T_{th}-t+2}\sum_{i=1}^tr^{2(i-1)}. \ene

\noindent Thus, the total data size at the end of the client recruitment phase can be solved as in (\ref{equ_BTth}). According to (\ref{equ_pt_lambda}) and (\ref{equ_BwithB(T)}), the optimal dynamic pricing $p(t)$ is obtained as in (\ref{equ_p(t)withoutB(t)}). Note that $\frac{b^3\tau^3D^2}{16\alpha^3s(\sum_{i=1}^{T_{th}}r^{2(i-1)})^3}$ will not change with $t$ for any given $T_{th}$. When $r<1$, $r^{5T_{th}-5t-6}$ increases with $t$, and thus $p(t)$ increases with $t$.

In the following, we will show that $p(t)\leq b(T - T_{th})$ is satisfied when $t\leq T_{th}-2$. According to (\ref{equ_p(t)withoutB(t)}), $p(t)\leq b(T - T_{th})$ is equivalent to
\bee\label{equ_p<b_equivalent} 16b^2 \alpha^3(T-T_{th})^3\frac{s}{\tau}\frac{(\sum_{i=1}^{T_{th}}r^{2(i-1)})^3}{r^{5T_{th}-5t-6}}\geq 1. \ene
For $r<1$, $\frac{(\sum_{i=1}^{T_{th}}r^{2(i-1)})^3}{r^{5T_{th}-5t-6}}>\frac{1}{r^{5T_{th}-5t-6}}>1$ always holds when $t\leq T_{th}-2$. Note that $b\geq 1, \frac{s}{\tau}\geq 1$, and $\alpha\geq0.5$. Thus, (\ref{equ_p<b_equivalent}) always holds when $t\leq T_{th}-2$.

When $t=T_{th}-1$, $p(T_{th}-1)\leq b(T - T_{th})$ holds if
\bee (1-r^2)^3\leq 2r(1-r^{2T_{th}})^3,\ene
which always holds if
\bee (1-r^2)^3\leq 2r(1-r^{2})^3, \ene
i.e., $r\geq 0.5$. Therefore, we can conclude that $p(t)\leq b(T - T_{th})$ is always satisfied when $t\leq T_{th}-1$. \qed


Proposition \ref{pro_dynamicpricing} shows that when time slot $t$ approaches the recruitment deadline $T_{th}$ or the data aging factor $r$ is large, it is necessary to increase the price to ensure recruiting enough data to train the model. 

\section{Optimal Recruitment Threshold in Stage~I}\label{sec_optThreshold}

Under the optimal dynamic pricing in Proposition \ref{pro_dynamicpricing}, a longer client recruitment time $T_{th}$ results in a larger total data size $B(T_{th})$ in (\ref{equ_BTth}) at the cost of less training iteration number $D$. Therefore, in Stage I the central server should find the optimal recruitment threshold $T_{th}$ to balance the total data size and training time for cost minimization in finite time horizon $T$, i.e.,
\bee T_{th}^*=\arg\min_{T_{th}\in\{1,...,T-1\}}U(T), \ene
where the total expected costs $U(T)$ under the optimal dynamic pricing $p(t)$ in (\ref{equ_p(t)withoutB(t)}) is:
\bee\begin{split}\label{equ_UT22} U(T)=&(4^{-\frac{4}{5}}+4^{\frac{1}{5}})(\frac{b\tau}{\alpha s^2r^2})^{\frac{1}{5}}(\frac{1-r^2}{1-r^{2T_{th}}})^{\frac{1}{5}}(\frac{\tau}{T-T_{th}})^{\frac{1}{5}}\\
&+\frac{\tau}{T-T_{th}}. \end{split}\ene


In the following proposition, the optimal threshold $T_{th}$ is derived in closed-form.
\begin{pro}
\label{thm_threshold} The optimal threshold $T_{th}^*\in\bZ^+$ depends on the training time $\tau$ per iteration and is given as follows:
\begin{itemize}
  \item Given high training time per iteration ($\tau \geq \overline{\psi}^{\frac{5}{3}}$), the server decides $T_{th}^*=1$ by recruiting clients in one time slot only and save more time for model training.
  \item Given low training time per iteration ($0<\tau<\overline{\psi}^{\frac{5}{3}}$), the server decides $T_{th}^*=\arg\min_{T_{th}}(U(T)|_{T_{th}=\lfloor\bar{T}_{th}\rfloor},U(T)|_{T_{th}=\lfloor\bar{T}_{th}\rfloor+1})\in\{1,\dots,T-1\}$, with $\bar{T}_{th}$ as the unique solution to
      \bee\label{equ_opt_Tth_for_SingleGroup}\begin{split} &\frac{1}{5}(4^{-\frac{4}{5}}+4^{\frac{1}{5}})(\frac{b\tau^2(1-r^2)}{\alpha s^2r^2(1-r^{2T_{th}})(T-T_{th})})^{\frac{1}{5}}\\
      \times&\Big(\frac{2r^{2T_{th}}\ln(r)}{1-r^{2T_{th}}}+\frac{1}{T-T_{th}}\Big)+\frac{\tau}{(T-T_{th})^2}=0, \end{split}\ene
\end{itemize}
where \bee\begin{split}\label{equ_overlinepsi} \overline{\psi}=&
\frac{1}{5}(\frac{b}{\alpha s^2r^2})^{\frac{1}{5}}(4^{-\frac{4}{5}}+4^{\frac{1}{5}})\times\\
&\Big(2|\ln(r)|\frac{r^{2}}{1-r^2}(T-1)^{\frac{9}{5}}-(T-1)^{\frac{4}{5}}\Big). \end{split}\ene
\end{pro}




\textbf{Proof:} Take the first-order derivative of (\ref{equ_UT22}) with respect to $T_{th}$, we have
\bee\label{equ_partialU(T)}\begin{split} \frac{\partial U(T)}{\partial T_{th}}=&\frac{1}{5}(4^{-\frac{4}{5}}+4^{\frac{1}{5}})(\frac{b\tau^2(1-r^2)}{\alpha s^2r^2(1-r^{2T_{th}})(T-T_{th})})^{\frac{1}{5}}\\
&\times\Big(\frac{2r^{2T_{th}}\ln(r)}{1-r^{2T_{th}}}+\frac{1}{T-T_{th}}\Big)+\frac{\tau}{(T-T_{th})^2}. \end{split}\ene

Since $\frac{\partial^2 U(T)}{\partial T_{th}^2}>0$ for any $0.5\leq r<1$, $T_{th}^*$ can be obtained according to $\frac{\partial U(T)}{\partial T_{th}}=0$.
Note that $\frac{\partial U(T)}{\partial T_{th}}$ increases in $T_{th}$. Therefore, we consider the following three cases:

(i) if $\frac{\partial U(T)}{\partial T_{th}}|_{T_{th}=1}\geq 0$, i.e.,
\bee\begin{split} \tau^{\frac{3}{5}}\geq & \frac{2}{5}(\frac{b}{\alpha s^2r^2})^{\frac{1}{5}}(4^{-\frac{4}{5}}+4^{\frac{1}{5}})|\ln(r)|\frac{r^{2}}{1-r^2}(T-1)^{\frac{9}{5}}\\
&-\frac{1}{5}(\frac{b}{\alpha s^2r^2})^{\frac{1}{5}}(4^{-\frac{4}{5}}+4^{\frac{1}{5}})(T-1)^{\frac{4}{5}}:=\overline{\psi}, \end{split}\ene
$\frac{\partial U(T)}{\partial T_{th}}\geq 0$ always holds for $T_{th}\in[1,T-1]$, which means $T_{th}^*=1$.

(ii) If $\frac{\partial U(T)}{\partial T_{th}}|_{T_{th}=T-1}\leq 0$, i.e.,
\bee\label{equ_psiunder}\begin{split} \tau^{\frac{3}{5}}\leq &\frac{2}{5}(\frac{b}{\alpha s^2r^2})^{\frac{1}{5}}(4^{-\frac{4}{5}}+4^{\frac{1}{5}})|\ln(r)|\frac{r^{2(T-1)}(1-r^2)^{\frac{1}{5}}}{(1-r^{2(T-1)})^{\frac{6}{5}}}\\
&-\frac{1}{5}(\frac{b}{\alpha s^2r^2})^{\frac{1}{5}}(4^{-\frac{4}{5}}+4^{\frac{1}{5}})(\frac{1-r^2}{1-r^{2(T-1)}})^{\frac{1}{5}}:=\underline{\psi}, \end{split}\ene
$\frac{\partial U(T)}{\partial T_{th}}\leq 0$ always holds for $T_{th}\in[1,T-1]$, which means $T_{th}^*=T-1$.

However, in the following, we can show that $\underline{\psi}$ is always negative for $T\geq 2$ and $r<1$. Thus, $\tau^{\frac{3}{5}}\leq \underline{\psi}$ doesn't exist.

According to (\ref{equ_psiunder}), $\underline{\psi}\leq 0$ is equivalent to
\bee\label{equ_prove(ii)} (1+2|\ln(r)|)r^{2(T-1)}\leq 1. \ene
Note that $\frac{\partial ((1+2|\ln(r)|)r^{2(T-1)})}{\partial r}>0$ always holds for $T\geq 2$ and $r<1$, and $\lim_{r\rightarrow 1}(1+2|\ln(r)|)r^{2(T-1)}=1$. Thus, (\ref{equ_prove(ii)}) always holds, i.e., $\underline{\psi}\leq 0$.

(iii) if $\frac{\partial U(T)}{\partial T_{th}}|_{T_{th}=1}<0$ and $\frac{\partial U(T)}{\partial T_{th}}|_{T_{th}=T-1}>0$, i.e., $0<\tau^{\frac{3}{5}}<\overline{\psi}$, since $\frac{\partial U(T)}{\partial T_{th}}$ increases in $T_{th}$, $T_{th}^*\in[1,\dots,T-1]$ can be obtained as the unique solution to $\frac{\partial U(T)}{\partial T_{th}}=0$. Note that here $T_{th}^*$ can still be equal to $1$ as the solution to $\frac{\partial U(T)}{\partial T_{th}}=0$ can be within the range $(1,2)$. Thus, the integer $T_{th}^*$ will be decided by comparing $U(T)|_{T_{th}=1}$ and $U(T)|_{T_{th}=2}$. \qed




As shown in Proposition \ref{thm_threshold}, if training time is high ($\tau \geq \overline{\psi}^{\frac{5}{3}}$), the FL training needs sufficient time to converge and thus we decide the minimum recruitment time $T_{th}^*=1$. As this requirement relaxes, we gradually increase the recruitment time $T_{th}^*$. Note that we obtain an expression for time partition $T_{th}^*$ in closed-form, and do not face any issue for computational complexity.

\begin{figure}
\centering\includegraphics[scale=0.3]{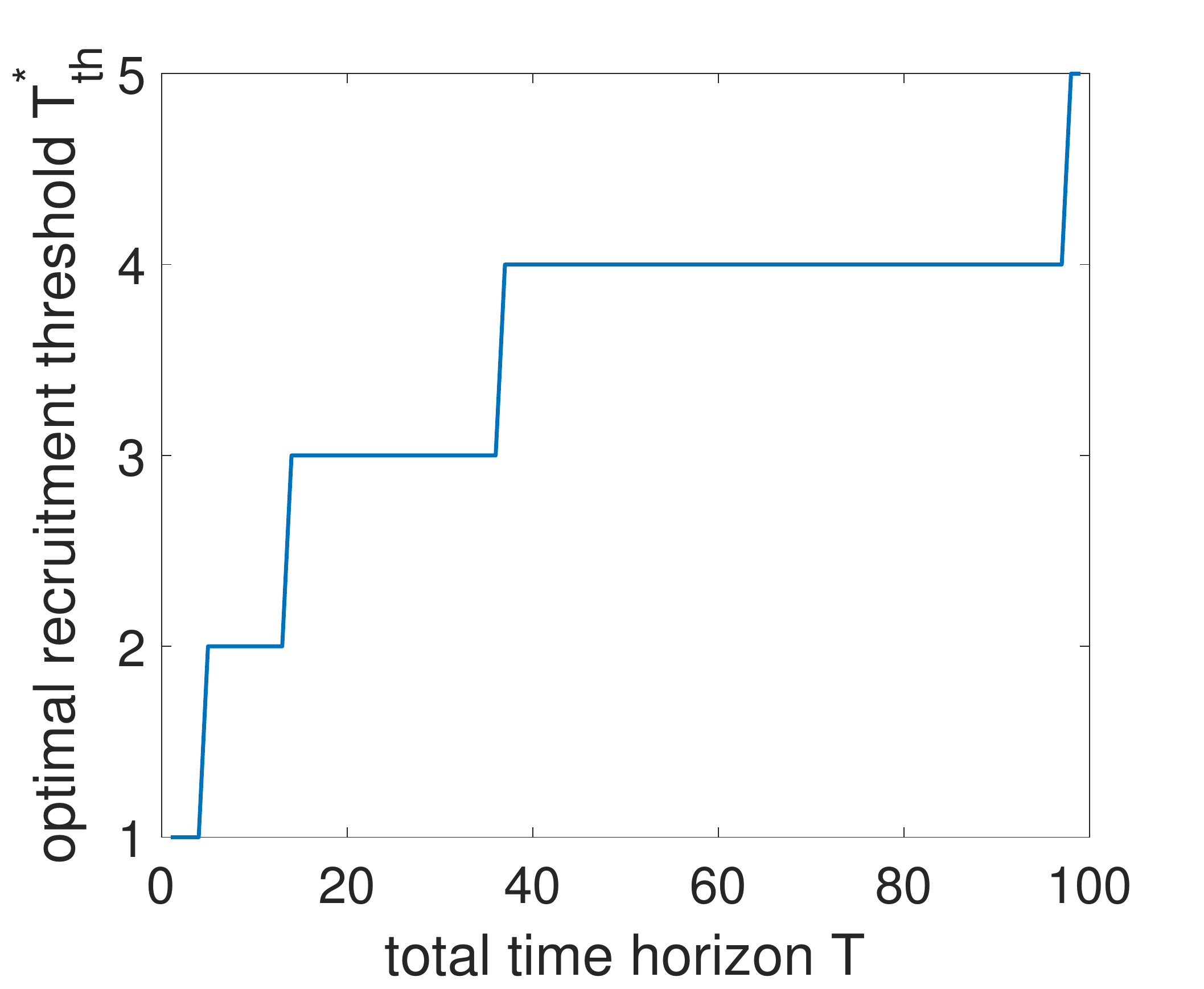}\caption{Optimal recruitment threshold $T_{th}^*$ versus total time horizon $T$.}\label{fig_TthVST}
\end{figure}

\begin{figure}
\centering\includegraphics[scale=0.28]{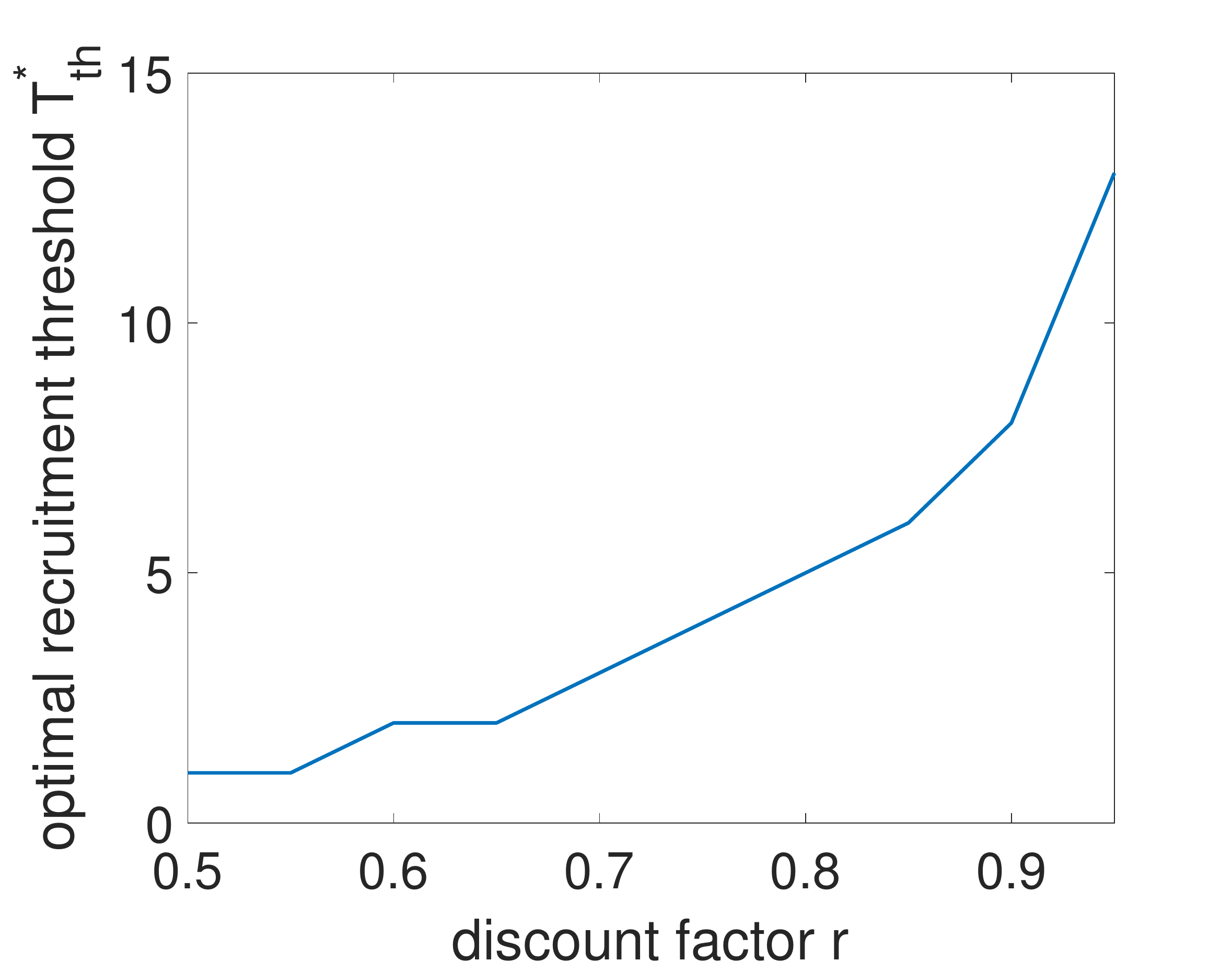}\caption{Optimal recruitment threshold $T_{th}^*$ versus discount factor $r$ when $T=50$.}\label{fig_TthVSr}
\end{figure}

To more clearly illustrate our results,  Fig. \ref{fig_TthVST} numerically shows that the optimal recruitment threshold $T_{th}^*$ increases with the total time horizon $T$ to relax the recruitment deadline. Moreover, Fig. \ref{fig_TthVSr} shows that the optimal recruitment threshold $T_{th}^*$ increases with the discount factor $r$, which tells the data aging effect in (\ref{equ_BUniform}). As the aging effect becomes weaker with greater $r$, the clients who arrived early in the recruitment phase still contribute a lot to the training dataset and we prolong the recruitment time $T_{th}^*$ to accommodate more useful data.





\section{Extension to Heterogeneous Clients}\label{sec_mutiType}

In previous sections, we have analyzed the optimal dynamic pricing and recruitment threshold for homogeneous clients with the identical data size and training time. In this section, we will extend to $N$ types of heterogeneous clients with different data sizes $s_i$ and training time $\tau_i, i\in\{1,...,N\}$. A client type may have multiple clients and will incur a longer iteration duration if provided with more data to compute \cite{tran2019federated}. Without loss of generality, we assume $(s_i,\tau_i), i \in \{1,...,N\}$ are sorted in ascending order, i.e., $s_1<s_2<\cdots<s_N, \tau_1<\tau_2<\cdots<\tau_N$. A client of type $i$ with private unit cost $c$ per training time will accept the price $p_i(t)$ at time slot $t$ if the price offer can well compensate his cost with $c\tau_iD\leq p_i(t),i=\{1,...,N\}$, where the training iteration number $D$ now varies according to our selection of client types for FL and $\tau_iD$ is the total training time of type-$i$ clients. In the synchronous FL setting, we set each iteration's duration to be the longest client computing time to wait for all the selected clients' updates.
For example, if the central server recruits clients of types $i\in\{1,2,3\}$, the global iteration number is $D=\frac{T-T_{th}}{\tau_3}$, where the clients of types $1$ and $2$ wait for the type-$3$ clients' updates.

Given such latency in model training, not necessarily all types of clients will be invited to help train the global model. To include clients with larger amounts of data, the iteration duration time is also longer, and the left iteration number becomes smaller in the model training phase. Thus, it is necessary to select appropriate client types to balance the data size and global iteration number. However, the optimal client-type choosing is a combinatorial optimization problem that has very high computational complexity. The client-type choice will affect the following recruitment threshold as well as the dynamic price for each client type. The joint optimization of them is NP-hard. The time complexity to only check every combination of the client-type choice is $O(2^{N})$ by increasing exponentially in $N$, not to mention the huge number of price combinations over time for each type-$i$ clients, with computation complexity $O((b\tau_iD/\epsilon)^{T_{th}})$ increasing exponentially in $T_{th}$. By considering the above issues, we formulate the decision process of the central server as the following three stages:
\begin{itemize}
  \item Stage I: The central server chooses the types of clients to recruit for model training.
  \item Stage II: Given the types of inviting clients, the central server decides the optimal recruitment threshold $T_{th}$.
  \item Stage III: Given the optimal recruitment threshold $T_{th}$ and inviting types of clients, the central server decides the dynamic pricing $p_i(t),t\in\{0,\dots,T_{th}-1\}$ for each inviting type-$i$ clients.
\end{itemize}

In the following, we use backward induction to analyze the above three-stage decision problem.

\subsection{Optimal Dynamic Pricing in Stage III}\label{sec_multiDP}

In this section, given the inviting types of client in Stage I and recruitment threshold $T_{th}$ in Stage II, we will analyze the optimal dynamic pricing $p_i(t),t\in\{0,\dots,T_{th}-1\}$ for each type-$i$ clients. Without loss of generality, we denote the inviting types of clients as $\{1,...,\bar{N}\}$ with $\bar{N}\in\{1,...,N\}$, and later in Section \ref{sec_usertype}, we prove the monotonic selection of client types. Denote the percentage of type-$i$ clients as $q_{i}$ with $\sum_{i=1}^{\bar{N}}q_i=1$. For uniform distribution of the clients' private costs, the probability that a type-$i$ client appears and accepts the price offer $p_i(t)$ at time $t$ is $\alpha q_i F(\frac{p_i(t)}{\tau_iD})=\frac{\alpha q_i}{b\tau_iD}p_i(t)$. By noting the data size $s_i$ contributed by type-$i$ client, the expected increase of data size at time $t$ is $\sum_{i=1}^{\bar{N}}\frac{\alpha q_i s_i p_i(t)}{b\tau_i D}$. Starting from the initial training data size $B(t=0)=0$, the dynamics of the expected data size $B(t)$ is given as (\ref{equ_BJ_update}).


Since the expected payment to type-$i$ client at time $t$ is $\frac{\alpha q_i}{b\tau_iD}p_i^2(t)$, the total expected payment at time $t$ is $\sum_{i=1}^{\bar{N}}\frac{\alpha q_i}{b\tau_iD}p_i^2(t)$. Note that the optimal price $p_i(t)$ for each type-$i$ clients should not exceed the maximum cost $b\tau_iD$ of this client-type as it is unnecessary for the provider to over-pay. Therefore, given the inviting types of clients $\{1,...,\bar{N}\}$ and recruitment threshold $T_{th}$, the central server aims to find the optimal dynamic pricing $p_i(t),t\in\{0,\dots,T_{th}-1\}$ for each type-$i$ clients to minimize the total expected cost $J(\boldsymbol{P(t)}$ $|T_{th}, \{1, ..., \bar{N}\})$ consisting of the total payment to the clients and the expected accuracy loss for FL. That is,

\bee\label{equ_J(T)}\begin{split} &J(\boldsymbol{P(t)}|T_{th}, \{1, ..., \bar{N}\})\\
=&\min_{
\substack{p_i(t)\leq b\tau_iD \\ t\in\{0,\dots,T_{th}-1\}}
}
\sum_{t=0}^{T_{th}-1}\sum_{i=1}^{\bar{N}}\frac{\alpha q_i}{b\tau_iD}p_i^2(t)+\frac{1}{\sqrt{B(T_{th})D}}+\frac{1}{D}\\
=&\min_{
\substack{p_i(t)\leq b\tau_iD\\ t\in\{0,\dots,T_{th}-1\}}
}\sum_{t=0}^{T_{th}-1}\Big(\boldsymbol{P(t)}^{\top}\boldsymbol{WP(t)}\Big)+\frac{1}{\sqrt{B(T_{th})D}}+\frac{1}{D}, \end{split}\ene
s.t. \bee\label{equ_BJ_update}\begin{split} B(t+1)=&r\Big(B(t)+\sum_{i=1}^{\bar{N}}\frac{\alpha q_i s_i p_i(t)}{b\tau_i D}\Big)=rB(t)+\boldsymbol{Q}\boldsymbol{P(t)}, \end{split}\ene
where
\begin{equation*}
\begin{aligned}
\boldsymbol{P(t)}=[p_1(t),...,p_{\bar{N}}(t)]^{\top}\in\bR^{\bar{N}\times 1} \\
\boldsymbol{Q}=[\frac{r\alpha q_1 s_1}{b\tau_1D},...,\frac{r\alpha q_{\bar{N}} s_{\bar{N}}}{b\tau_{\bar{N}}D}]\in\bR^{1\times {\bar{N}}} \\
\boldsymbol{W}=\begin{bmatrix}
   \frac{\alpha q_1}{b\tau_1D}  & \cdots\ & 0\\
\vdots & \ddots  & \vdots  \\
 0  & \cdots\ & \frac{\alpha q_{\bar{N}}}{b\tau_{\bar{N}}D}\\
  \end{bmatrix}\in\bR^{\bar{N} \times \bar{N}}.
\end{aligned}
\end{equation*}

Similar to the analysis of Proposition \ref{pro_dynamicpricing}, we have the following lemma.

\begin{lem}\label{opt_dp_single} The optimal dynamic pricing
$\boldsymbol{P(t)} \in \bR^{\bar{N} \times 1}$ is
\bee\label{equ_pt_multi}\begin{split} \boldsymbol{P(t)} = min(
   \begin{bmatrix}
   s_{1} \Gamma_t \\
   \vdots \\
   s_{\bar{N}} \Gamma_t
   \end{bmatrix},
   \begin{bmatrix}
   b\tau_1D \\
   \vdots \\
   b\tau_{\bar{N}}D
   \end{bmatrix}
  ).
  \end{split}
\ene
where $\Gamma_t=\big(\frac{b^3D^2r^{5T_{th}-5t-6}(1-r^{2})^3}{16\alpha^3(1-r^{2T_{th}})^3 (\sum_{i=1}^{\bar{N}}\frac{q_is_i^2}{\tau_i})^3}\big)^{\frac{1}{5}}$, which is monotonically increasing in $t$.\footnote{If $s_j\Gamma_t>b\tau_jD$ for client type $j$ from certain time slot $t'$, the optimal pricing is $p_j(t)=b\tau_jD$ for any $t\geq t'$.}
\par

\end{lem}



 \textbf{Proof:} According to the problem (\ref{equ_J(T)})-(\ref{equ_BJ_update}), we can construct the Hamiltonian function as follows:
 \bee H(t)=\boldsymbol{P(t)}^{\top}\boldsymbol{WP(t)}+\lambda(t+1)((r-1)B(t)+\boldsymbol{Q}\boldsymbol{P(t)}).\ene

Since $\frac{\partial^2 H(t)}{\partial p_i^2(t)}>0, i\in\{1,...,\bar{N}\}$, the Hamiltonian function is convex in $p_i(t)$. Similar to the proof of Proposition \ref{pro_dynamicpricing}, we have

 \bee\label{equ_P(t)m} \boldsymbol{P(t)}=\frac{1}{4}\boldsymbol{W}^{-1}\boldsymbol{Q}^{\top}D^{-\frac{1}{2}}(B(T_{th}))^{-\frac{3}{2}}r^{T_{th}-t-1}, \ene
 and
 \bee\label{equ_B(t)m} B(t)=\frac{1}{4}\boldsymbol{Q}\boldsymbol{W}^{-1}\boldsymbol{Q}^{\top}D^{-\frac{1}{2}}(B(T_{th}))^{-\frac{3}{2}}r^{T_{th}-t}\sum_{i=1}^tr^{2(i-1)}. \ene

 According to (\ref{equ_B(t)m}), $B(T_{th})$ is solved as:
 \bee\label{equ_BTthm} B(T_{th})=\Big(\frac{1}{4}\boldsymbol{Q}\boldsymbol{W}^{-1}\boldsymbol{Q}^{\top}D^{-\frac{1}{2}}\sum_{i=1}^{T_{th}}r^{2(i-1)}\Big)^{\frac{2}{5}}, \ene
 Insert (\ref{equ_BTthm}) into (\ref{equ_P(t)m}) and note that $p_i(t)\leq b\tau_iD$ for any $i\in\{1,...,\bar{N}\}$, the optimal dynamic pricing $P(t)$ can be derived as (\ref{equ_pt_multi}). Since $r^{5T_{th}-5t-6}$ increases with $t$ for any $0<r<1$, $P(t)$ is monotonically increasing with $t$.

 Note that the Hamiltonian function is convex in $p_i(t), i\in\{1,...,\bar{N}\}$ for the problem (\ref{equ_J(T)})-(\ref{equ_BJ_update}). Thus, when $p_i(t)> b\tau_iD$, it's optimal to set $p_i(t)=b\tau_iD$. Moreover, by noting that $p_i(t), i\in\{1,...,\bar{N}\}$ increases with $t$, if $\Gamma_t s_j>b\tau_jD$ for client type $j$ from certain time slot $t'$, the optimal pricing is $p_j(t)=b\tau_jD$ for any $t\geq t'$. \qed



According to Lemma \ref{opt_dp_single}, if $ s_{i} \Gamma_t\leq b\tau_iD$ holds for any client-type $i\in\{1,..,\bar{N}\}$, the resulting total expected cost is:
      \bee\label{equ_JTtotal}\begin{split} &J_{\leq}(\boldsymbol{P(t)}|T_{th},\{1, ..., \bar{N}\})\\
      =&(4^{-\frac{4}{5}}+4^{\frac{1}{5}})(\frac{b}{\alpha r^2})^{\frac{1}{5}}(\frac{1-r^2}{1-r^{2T_{th}}})^{\frac{1}{5}}(\sum_{i=1}^{\bar{N}}\frac{q_is_i^2}{\tau_i})^{-\frac{1}{5}}(\frac{\tau_{\bar{N}}}{T-T_{th}})^{\frac{1}{5}}\\
      &+\frac{\tau_{\bar{N}}}{T-T_{th}}. \end{split}
      \ene
Lemma \ref{opt_dp_single} also shows that for the client type with large amount of data, higher dynamic prices are required to compensate for their higher training costs.  Also, like homogeneous clients, the dynamic price increases over time due to data aging (that is, the value of data decreases over time).

\subsection{Optimal Recruitment Threshold in Stage II}

Based on the optimal dynamic pricing in Section \ref{sec_multiDP}, in Stage II we are ready to analyze the optimal recruitment threshold $T_{th}$ given the types $\{1,...,\bar{N}\}$ of inviting clients in Stage I. We propose the following proposition to find the optimal threshold $T_{th}$ for heterogeneous clients case.

\begin{pro}
\label{thm_threshold_for_multi} The optimal threshold $T_{th}^*\in\bZ^+$ for heterogeneous clients depends on their training rate $\frac{s_{i}}{\tau_{i}}$ (data size/training time) and is given as follows:
\begin{itemize}
  \item Given low training rate ($\frac{s_{i}}{\tau_{i}}  \leq  \frac{bD}{\Gamma_t'}$ for any $i \in \{1, ..., \bar{N}\}, t \in \{0, ..., \tilde{T}_{th} - 1\}$), the server decides $T_{th}^*=\tilde{T}_{th}$.

  \item  Given high training rate ( $\frac{s_{i}}{\tau_{i}}  >  \frac{bD}{\Gamma_t'}$ for certain $i \in \{1, ..., \bar{N}\}, t \in \{0, ..., \tilde{T}_{th} - 1\}$), the optimal recruitment threshold $T_{th}^*$ can be obtained according to Algorithm \ref{alg_optimalTthMultiple}.
\end{itemize}
where $\Gamma_t'=\big(\frac{b^3D^2r^{5\tilde{T}_{th}-5t-6}(1-r^{2})^3}{16\alpha^3(1-r^{2\tilde{T}_{th}})^3 (\sum_{i=1}^{\bar{N}}\frac{q_is_i^2}{\tau_i})^3}\big)^{\frac{1}{5}}$ and $\tilde{T}_{th}=\arg\min_{T_{th}}($ $J_{\leq}(\boldsymbol{P(t)}|\lfloor\hat{T}_{th}\rfloor, \{1, ..., \bar{N}\}),J_{\leq}(\boldsymbol{P(t)}|\lfloor\hat{T}_{th}\rfloor+1, \{1, ..., \bar{N}\}))\in\{1,...,T-1\}$ with $\hat{T}_{th}\geq 1$ as the unique solution to

      \bee\label{equ_partialJ(T)}\begin{split} &\frac{1}{5}(4^{-\frac{4}{5}}+4^{\frac{1}{5}})(\sum_{i=1}^{\bar{N}}\frac{q_is_i^2}{\tau_i})^{-\frac{1}{5}}(\frac{b\tau_{\bar{N}}(1-r^2)}{\alpha r^2(1-r^{2T_{th}})(T-T_{th})})^{\frac{1}{5}}\\
&\times\Big(\frac{2r^{2T_{th}}\ln(r)}{1-r^{2T_{th}}}+\frac{1}{T-T_{th}}\Big)+\frac{\tau}{(T-T_{th})^2}=0. \end{split}\ene

\end{pro}


 \textbf{Proof:} Based on Lemma \ref{opt_dp_single}, we consider the following two cases based on whether $\Gamma_t s_i\leq b\tau_iD$:

(i) First, we consider the case that $\Gamma_t s_i\leq b\tau_iD$ is always satisfied for any $i\in\{1,...,\bar{N}\}$, $t\in\{0,\dots,T_{th}-1\}$. According to $\boldsymbol{P(t)}$ in (\ref{equ_pt_multi}), we obtain the total expected cost $J_{\leq}(\boldsymbol{P(t)}|T_{th},\{1,...,\bar{N}\})$ in (\ref{equ_JTtotal}). We can check that $\frac{\partial^2 J_{\leq}(\boldsymbol{P(t)}|T_{th}, \{1,...,\bar{N}\})}{\partial T_{th}^2} > 0$ is always satisfied, and thus the optimal $T_{th}^*$ can be obtained based on the first-order conditions $\frac{\partial J_{\leq}(\boldsymbol{P(t)} |T_{th}, \{1,...,\bar{N}\})}{\partial T_{th}}=0$, which is given in (\ref{equ_partialJ(T)}).

 Denote the solution to (\ref{equ_partialJ(T)}) as $\hat{T}_{th}$. Note that $\frac{\partial J_{\leq}(\boldsymbol{P(t)} |T_{th}, \{1,...,\bar{N}\})}{\partial T_{th}}$ $=0$ increases in $T_{th}$. If $\frac{\partial J_{\leq}(\boldsymbol{P(t)}|T_{th},\{1, ..., \bar{N}\})}{\partial T_{th}}|_{T_{th}=1}>0$, we have $\hat{T}_{th}<1$. Since $J_{\leq}(\boldsymbol{P(t)}|T_{th},\{1, ..., \bar{N}\})$ is convex in $T_{th}$, the optimal threshold $T_{th}^*=\tilde{T}_{th}=\hat{T}_{th}=1$. If $\frac{\partial J_{\leq}(\boldsymbol{P(t)}|T_{th},\{1, ..., \bar{N}\})}{\partial T_{th}}|_{T_{th}=1}\leq 0$, $\hat{T}_{th}\geq 1$ is the unique solution to (\ref{equ_partialJ(T)}). By noting that $T_{th}\in\bZ^+$, the optimal threshold is $T_{th}^*=\tilde{T}_{th}=\arg\min_{T_{th}}(J_{\leq}(\boldsymbol{P(t)}|\lfloor\hat{T}_{th}\rfloor, \{1, ...,$ $\bar{N}\})$, $J_{\leq}(\boldsymbol{P(t)}|\lfloor\hat{T}_{th}\rfloor+1, \{1, ..., \bar{N}\}))\in\{1,...,T-1\}$. By noting that $\Gamma_t$ is a function of $T_{th}$, we define $\Gamma_t'=\big(\frac{b^3D^2r^{5\tilde{T}_{th}-5t-6}(1-r^{2})^3}{16\alpha^3(1-r^{2\tilde{T}_{th}})^3 (\sum_{i=1}^{\bar{N}}\frac{q_is_i^2}{\tau_i})^3}\big)^{\frac{1}{5}}$ given $\tilde{T}_{th}$.
 Therefore, we can conclude that, if $\frac{s_{i}}{\tau_{i}}  \leq  \frac{bD}{\Gamma_t'}$ for any $i \in \{1, ..., \bar{N}\}, t \in \{0, ..., \tilde{T}_{th} - 1\}$, $p_i(t)=\Gamma_t' s_i\leq b\tau_iD$ is always satisfied and the optimal threshold $T_{th}^*=\tilde{T}_{th}$.

(ii) Then, we consider the case that $\Gamma_t' s_{i} > b\tau_iD$ for certain client type $i\in\{1,...,\bar{N}\}$ at time $t'\in\{0,\dots,\tilde{T}_{th}-1\}$, the optimal pricing for client type $i$ is $p_i(t)=b\tau_iD$ for any $t\geq t'$. Based on the data size updating dynamics (\ref{equ_BJ_update}), the total batch size $B(T_{th})$ at time $T_{th}$ and the resulting total expected cost $J(\boldsymbol{P(t)}|T_{th},\{1, ..., j\})$ in (\ref{equ_J(T)}) can be derived given the optimal dynamic pricing $\boldsymbol{P(t)}$ in (\ref{equ_pt_multi}). Then, similar to the above analysis, the optimal recruitment threshold $T_{th}^*$ is obtained by checking the first-order condition as shown in Algorithm~\ref{alg_optimalTthMultiple}.
 \qed




The procedure to find the optimal recruitment threshold $T_{th}^*$ for heterogeneous clients is concluded in linear Algorithm~\ref{alg_optimalTthMultiple}. Note that if the solution to (\ref{equ_partialJ(T)}) is less than 1, $\hat{T}_{th}=1$, and then we can calculate $\tilde{T}_{th}$ and $\Gamma_t'$ to
check whether $\frac{s_{i}}{\tau_{i}} \leq \frac{bD}{\Gamma_t'}$ is satisfied for any $i \in \{1, .., \bar{N}\}, t \in \{0, ..., T_{th} - 1\}$. If satisfied, $T_{th}^*=\tilde{T}_{th}$. Otherwise, the total expected cost $J(\boldsymbol{P(t)}|T_{th},\{1, ..., \bar{N}\})$ can be calculated according to the dynamics of data size $B(t)$ in (\ref{equ_BJ_update}) and optimal dynamic pricing $P(t)$ in (\ref{equ_pt_multi}), and then the optimal recruitment threshold $T_{th}^*$ can be derived by checking the first-order condition $\frac{\partial J(\boldsymbol{P(t)}|T_{th},\{1, ..., \bar{N}\})}{\partial T_{th}}$. 

\begin{algorithm}[t]
\caption{Optimal recruitment threshold $T_{th}^*$ and client-type choice for heterogeneous clients.}
\begin{algorithmic}[1]

\FOR {$j=1:N$}

\STATE Solve $\hat{T}_{th}$ as the solution to (\ref{equ_partialJ(T)})
\IF {$\hat{T}_{th}<1$}
\STATE $\hat{T}_{th}=1$
\ENDIF
\STATE Calculate $\tilde{T}_{th}=\arg\min_{T_{th}}(J_{\leq}(\boldsymbol{P(t)}|\lfloor\hat{T}_{th}\rfloor, \{1, ..., \bar{N}\}),$
\STATE ~~~~~$J_{\leq}(\boldsymbol{P(t)}|\lfloor\hat{T}_{th}\rfloor+1, \{1, ..., \bar{N}\}))$ and $\Gamma_t'$
\IF {$\forall i \in \{1, ..., j\}, t \in \{0, ..., T_{th} - 1\}, \frac{s_i}{\tau_{i}} \leq  \frac{bD}{\Gamma_t'}$}
\RETURN $T_{th}^*=\tilde{T}_{th}$ and \\$J^*(\boldsymbol{P(t)}|T_{th}^*,\{1, ..., j\})=J_{\leq}(\boldsymbol{P(t)}|T_{th}^*,\{1, ..., j\})$
\ELSE
\STATE Calculate $J(\boldsymbol{P(t)}|T_{th},\{1, ..., j\})$ according to the dynamics of data size $B(t)$ in (\ref{equ_BJ_update}) and optimal dynamic pricing $P(t)$ in (\ref{equ_pt_multi})

\IF {$\frac{\partial J(\boldsymbol{P(t)}|T_{th},\{1, ..., j\})}{\partial T_{th}}|_{T_{th}=1}\geq 0$}
\RETURN $T_{th}^*=1$ and $J^*(\boldsymbol{P(t)}|T_{th}^*,\{1, ..., j\})$
\ELSE
\STATE Solve $\frac{\partial J(\boldsymbol{P(t)}|T_{th},\{1, ..., j\})}{\partial T_{th}}=0$ 
\RETURN $T_{th}^*=\arg\min(J(\boldsymbol{P(t)} | \lfloor T_{th}\rfloor, \{1,...,j\}),$ $J(\boldsymbol{P(t)} | \lfloor T_{th}\rfloor+1, \{1,...,j\}))$ and $J^*(\boldsymbol{P(t)}|T_{th}^*,\{1, ..., j\})$
\ENDIF
\ENDIF

\ENDFOR

\STATE $j^*=\arg\min(J^*(\boldsymbol{P(t)}|T_{th}^*,\{1, ..., j\})|j=1,...,N)$
\RETURN optimal client-type $\{1,2,...,j^*\}$

\end{algorithmic}
\label{alg_optimalTthMultiple}
\end{algorithm}

\subsection{Optimal Client-Type Choice in Stage I}\label{sec_usertype}

In this subsection, we will discuss the optimal types of clients to be invited in Stage I based on the optimal dynamic pricing and optimal recruitment threshold obtained above. Even though we have derived closed-form solutions to the optimal dynamic pricing and optimal recruitment threshold, the time complexity of finding the optimal client-type choice is still very high with $O(2^{N})$ increasing exponentially in $N$. In the following sections, we consider the non-trivial case that $p_i(t)=s_{i}\Gamma_t \leq b\tau_iD$ for any $i\in\{1,...,N\}$ to reveal the analytical result in Proposition \ref{pro_optimal_user_type}. 


\begin{pro}\label{pro_optimal_user_type} For any $N$ types of clients at the optimum, the central server monotonically chooses client types in set  $\{1,2,...,j^*\}$ with $j^*=\arg\min_{j\in\{1,...,N\}}J^*(\boldsymbol{P(t)}|T_{th}^*,\{1, ..., j\})$.
\end{pro}


 \textbf{Proof:} According to (\ref{equ_JTtotal}), the total expected cost for inviting type $i$ clients only is
 \bee\label{equ_J_iT}\begin{split} &J(\boldsymbol{P(t)}|T_{th},\{i\})\\=&(4^{-\frac{4}{5}}+4^{\frac{1}{5}})(\frac{b\tau_i}{\alpha q_i s_i^2r^2})^{\frac{1}{5}}(\frac{1-r^2}{1-r^{2T_{th}}})^{\frac{1}{5}}(\frac{\tau_i}{T-T_{th}})^{\frac{1}{5}}\\&+\frac{\tau_i}{T-T_{th}}. \end{split}\ene


 When there are two types of clients, i.e., $N=2$, according to (\ref{equ_J_iT}) and (\ref{equ_JTtotal}), we have $J(\boldsymbol{P(t)}|T_{th}, \{2\})\geq J(\boldsymbol{P(t)}|T_{th}, \{1,2\})$ for any given $T_{th}$. Thus, inviting two types of clients $\{1,2\}$ is always better than only inviting type-$2$ clients.

 When there are three types of clients, i.e., $N=3$, the possible client-type combinations are $\{1\}, \{2\}, \{3\}, \{1,2\}, \{1,3\}, \{2,3\}$ and $\{1,2,3\}$. According to (\ref{equ_JTtotal}), the total expected cost of inviting client types $\{1,3\}$ is
 \bee\label{equ_JTtotal_1_3}\begin{split} J(\boldsymbol{P(t)}|T_{th},\{1,3\})=&(4^{-\frac{4}{5}}+4^{\frac{1}{5}})(\frac{b}{\alpha r^2})^{\frac{1}{5}}(\frac{1-r^2}{1-r^{2T_{th}}})^{\frac{1}{5}}\\
 \times(\frac{q_1s_1^2}{\tau_1}&+\frac{q_3s_3^2}{\tau_3})^{-\frac{1}{5}}(\frac{\tau_3}{T-T_{th}})^{\frac{1}{5}}+\frac{\tau_3}{T-T_{th}}, \end{split}\ene
 and the total expected cost of inviting client types $\{1,2,3\}$ is
 \bee\label{equ_JTtotal_1_2_3}\begin{split} J(\boldsymbol{P(t)}|T_{th},\{1,2,3\})=&(4^{-\frac{4}{5}}+4^{\frac{1}{5}})(\frac{b}{\alpha r^2})^{\frac{1}{5}}(\frac{1-r^2}{1-r^{2T_{th}}})^{\frac{1}{5}}\\
 \times(\sum_{i=1}^3&\frac{q_is_i^2}{\tau_i})^{-\frac{1}{5}}(\frac{\tau_3}{T-T_{th}})^{\frac{1}{5}}+\frac{\tau_3}{T-T_{th}}. \end{split}\ene

 Since $\frac{q_1s_1^2}{\tau_1}+\frac{q_2s_2^2}{\tau_2}+\frac{q_3s_3^2}{\tau_3}>\frac{q_1s_1^2}{\tau_1}+\frac{q_3s_3^2}{\tau_3}$, $J(\boldsymbol{P(t)}|T_{th}, \{1,2,3\})<J(\boldsymbol{P(t)}|T_{th},\{1,3\})$ for any given $T_{th}$. Similarly, we have $J(\boldsymbol{P(t)}|T_{th}$, $\{1,2,3\}) < J(\boldsymbol{P(t)} | T_{th},\{2,3\}) \text{ and } J(\boldsymbol{P(t)}|T_{th}, \{1,2,3\}) < J(\boldsymbol{P(t)}|$ $T_{th}, \{3\})$. Note that $J(\boldsymbol{P(t)}|T_{th}, \{2\})\geq J(\boldsymbol{P(t)}|T_{th}, \{1,2\})$. Thus, when $N=3$, we only need to compare the total expected costs $J(\boldsymbol{P(t)}|T_{th},\{1\}), J(\boldsymbol{P(t)}|T_{th}, \{1,2\}), J(\boldsymbol{P(t)}|T_{th}, \{1,2,3\})$ to find the optimal types of inviting clients.

 Similar to the above analysis, for any $N$ groups of clients, if client-type $j$ is invited, all client-types $i\leq j$ should be invited for cost minimization. Thus, the central server only need to compare the optimal expected costs $J^*(\boldsymbol{P(t)}|T_{th}^*,\{1, ..., j\}), j\in\{1,...,N\}$ given the optimal recruitment threshold $T_{th}^*$ and optimal dynamic pricing in (\ref{equ_pt_multi}). Thus, the optimal types of inviting clients is $\{1,2,...,$ $j^*\}$ with $j^*=\arg\min_{j\in\{1,...,N\}}J^*(\boldsymbol{P(t)}|T_{th}^*,\{1, ..., j\})$. \qed

Given a selected client type $j$, the iteration duration is at least $\tau_j$. Thus including any client type $i<j$ with smaller data size and training time only creates updates within the iteration duration without reducing the number of global training iterations. As shown in Proposition \ref{pro_optimal_user_type}, we monotonically select the first $j^*$ types of clients with smaller data sizes and training time to accelerate the model training. The optimal types of inviting clients is one of the following cases: $\{1\}, \{1,2\}$, $..., \{1,2,...,N\}$. This means the multiple client-type choice is monotonic, which can reduce the time complexity of finding the optimal client-type choice from $O(2^{N})$ to $O(N)$ by enumerating only $N$ subsets as shown in Algorithm \ref{alg_optimalTthMultiple}.


\subsection{Robustness to Data Size}

In reality, our estimation of each client type may not be precise due to some noises, and we wonder our solution's robustness against estimation error of clients’ data size. Assume the data size $s_i(t)$ contributed by a type-$i$ client at time slot $t$ faces a variable and bounded error from our estimation: $s_i(t)\in[s_i-\delta_i, s_i+\delta_i]$, $0<\delta_i<s_i$, where $s_i$ can be viewed as the mean of type-$i$ clients' data size. Given the optimal client type choice $\{1,...,j^{*}\}$, by applying the dynamic pricing $p_i(t)$ in (\ref{equ_pt_multi}), the dynamics of data size $B(t)$ under uncertain client data size $s_i(t), i\in \{1, ..., j^{*}\}$ changes from (\ref{equ_BJ_update}) to
\bee\label{equ_BJ_changing_s} \tilde{B}(t+1)=r\tilde{B}(t)+r\sum_{i=1}^{{j^{*}}}\frac{\alpha q_i s_i(t)}{b\tau_i D} p_i(t), \ene
and the resulting total expected cost is
\bee\label{equ_tildeJ(T)}\begin{split} \tilde{J}(\boldsymbol{P(t)}|T_{th}, \{1, ...,j^{*}\})=&\sum_{t=0}^{T_{th}-1}\sum_{i=1}^{j^{*}}\frac{\alpha q_i}{b\tau_iD}(p_i(t))^2\\
&+\frac{1}{\sqrt{\tilde{B}(T_{th})D}}+\frac{1}{D}. \end{split}\ene

By comparing the total expected cost $\tilde{J}(\boldsymbol{P(t)}|T_{th}, \{1,...,j^{*}\})$ under noisy data size $s_i(t)$ in (\ref{equ_tildeJ(T)}) with the total expected cost $J(\boldsymbol{P(t)}|$ $T_{th}, \{1,...,j^{*}\})$ under no noise case where all type-$i$ clients contribute the precisely data size $s_i$ in (\ref{equ_J(T)}), we have the following proposition.

\begin{pro}\label{pro_robust} By applying the optimal dynamic pricing $P(t)$ in (\ref{equ_pt_multi}), the total expected cost objective  $\tilde{J}(\boldsymbol{P(t)}|T_{th}, \{1,...,j^{*}\})$ with noisy data size $s_i(t)\in[s_i-\delta_i, s_i+\delta_i], i\in\{1,..., j^{*}\}, t\in\{0,...,T_{th}-1\}$ satisfy
\bee\label{equ_prorobust}\begin{split} \tilde{J}(\boldsymbol{P(t)}|T_{th}, \{1, ..., j^{*}\})\leq & J(\boldsymbol{P(t)}|T_{th}, \{1, ..., j^{*}\})\\
&+\Phi(\delta_i|i \in \{1, ..., j^{*}\}), \end{split}\ene
where \bee\begin{split} \Phi(\delta_i|i\in\{1,...,j^{*}\})&=\Big(\frac{4b\tau_{j^{*}}(1-r^2)}{\alpha r^2(T-T_{th})(1-r^{2T_{th}})}\Big)^{\frac{1}{5}}\\
\times\Big(&\frac{(\sum_{i=1}^{j^{*}}\frac{q_is_i^2}{\tau_i})^{\frac{3}{10}}}{(\sum_{i=1}^{j^{*}}\frac{q_is_i(s_i-\delta_i)}{\tau_i})^{\frac{1}{2}}}-(\sum_{i=1}^{j^{*}}\frac{q_is_i^2}{\tau_i})^{-\frac{1}{5}}\Big), \end{split}\ene
which increases with $\delta_i, i\in\{1,...,j^{*}\}$.
\end{pro}


 \textbf{Proof:} According to (\ref{equ_J(T)}), the total expected cost increases as the total data size $B(T_{th})$ decreases. Thus, given the optimal dynamic pricing $p_i(t)$ in (\ref{equ_pt_multi}) and recruitment threshold $T_{th}^*$, the worst-case total cost is:
 \bee\begin{split}
 &\bar{J}(\boldsymbol{P(t)}|T_{th}, \{1, ..., j^{*}\})\\=& \max_{
 \substack{s_i(t)\in[s_i-\delta_i, s_i+\delta_i], \\
 i\in\{1,...,j^{*}\}, \\
 t \in \{0,...,T_{th}-1\}}}
 \tilde{J}(\boldsymbol{P(t)}|T_{th}, \{1, ..., j^{*}\})
 \end{split}\ene
 which is achieved when $s_i(t)=s_i-\delta_i$ for any $i\in\{1,...,j^{*}\}, t\in\{0,...,T_{th}-1\}$. In this case, according to (\ref{equ_BJ_changing_s}) and $p_i(t)$ in (\ref{equ_pt_multi}), the total data size at time $T_{th}$ is

 \bee\label{equ_BTth_bar}\begin{split} \bar{B}(T_{th})=&\Big(\frac{\alpha^2}{16b^2D^3(\sum_{i=1}^{j^{*}}\frac{q_is_i^2}{\tau_i})^3}\Big)^{\frac{1}{5}}\sum_{i=1}^{j^{*}}\frac{q_is_i(s_i-\delta_i)}{\tau_i}\\
 &\times(\frac{r^2(1-r^{2T_{th}})}{1-r^2})^{\frac{2}{5}}.     \end{split}\ene

The worst-case total cost given $p_i(t)$ in (\ref{equ_pt_multi}) and $\bar{B}(T_{th})$ in (\ref{equ_BTth_bar}) is
 \bee\begin{split} &\bar{J}(\boldsymbol{P(t)}|T_{th}, \{1,...,j^{*}\})\\
 =&\frac{\tau_{j^{*}}}{T-T_{th}}+\Big(\frac{b\tau_{j^{*}} (1-r^2)}{\alpha r^2(T-T_{th})(1-r^{2T_{th}})}\Big)^{\frac{1}{5}}\\
 &\times\Big(4^{\frac{1}{5}}\frac{(\sum_{i=1}^{j^{*}}\frac{q_is_i^2}{\tau_i})^{\frac{3}{10}}}{(\sum_{i=1}^{j^{*}}\frac{q_is_i(s_i-\delta_i)}{\tau_i})^{\frac{1}{2}}}+4^{-\frac{4}{5}}(\sum_{i=1}^{j^{*}}\frac{q_is_i^2}{\tau_i})^{-\frac{1}{5}}\Big). \end{split}\ene

 By comparing the worst-case total cost $\bar{J}(\boldsymbol{P(t)}|T_{th}, \{1,...,j^{*}\})$ with the optimal total expected cost $J(\boldsymbol{P(t)}|T_{th}, \{1,...,j^{*}\})$ under no noise case where all type-$i$ clients contribute the precisely data size $s_i, i\in\{1,...$, $j^{*}\}$, we have
 \been\begin{split} &\Phi(\delta_i|i\in\{1,...,j^{*}\})\\=&\bar{J}(T|i\in\{1,...,j^{*}\})-J(T|i\in\{1,...,j^{*}\})\\
 =&\Big(\frac{4b\tau_{j^{*}}(1-r^2)}{\alpha r^2(T-T_{th})(1-r^{2T_{th}})}\Big)^{\frac{1}{5}}\\
 &\times\Big(\frac{(\sum_{i=1}^{j^{*}}\frac{q_is_i^2}{\tau_i})^{\frac{3}{10}}}{(\sum_{i=1}^{j^{*}}\frac{q_is_i(s_i-\delta_i)}{\tau_i})^{\frac{1}{2}}}-(\sum_{i=1}^{j^{*}}\frac{q_is_i^2}{\tau_i})^{-\frac{1}{5}}\Big), \end{split}\enen
 where $J(\boldsymbol{P(t)}|T_{th}, \{1,...,j^{*}\})$ is as given in (\ref{equ_JTtotal}). Note that $\tilde{J}(\boldsymbol{P(t)}|T_{th}$, $\{1,...,j^{*}\}) \leq \bar{J}(\boldsymbol{P(t)}|T_{th}, \{1,...,j^{*}\})$. Then, (\ref{equ_prorobust}) is obtained. \qed
 
 \begin{figure}
\centering\includegraphics[scale=0.38]{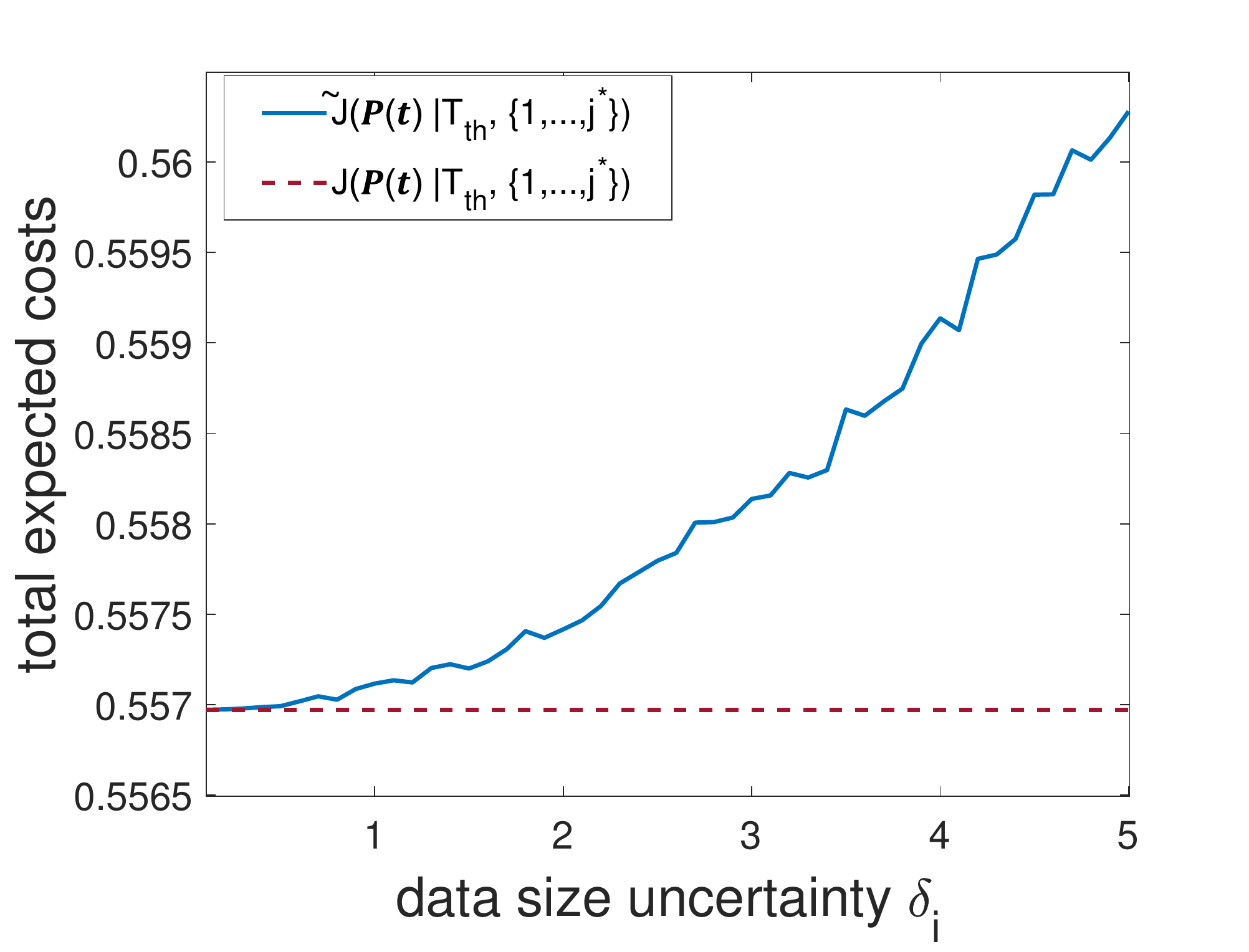}\caption{Total expected cost $\tilde{J}(\boldsymbol{P(t)}|T_{th}, \{1, ...,j^{*}\}\})$ versus data size 's noise bound $\delta_i$ for identical $\delta_i, i\in\{1,...,j^{*}\}$.}\label{fig_U_TVSDelta}
\end{figure}

\newcounter{myeq1}
\begin{figure*}[t]
\setcounter{myeq1}{\value{figure}}
\setcounter{figure}{5}
\centering
\subfigure[Optimal client-type choice $j^*$ versus data size disparity $\mu$.]{\label{fig_GroupswithDelta}
\begin{minipage}{.35\textwidth}
\includegraphics[width=1\textwidth]{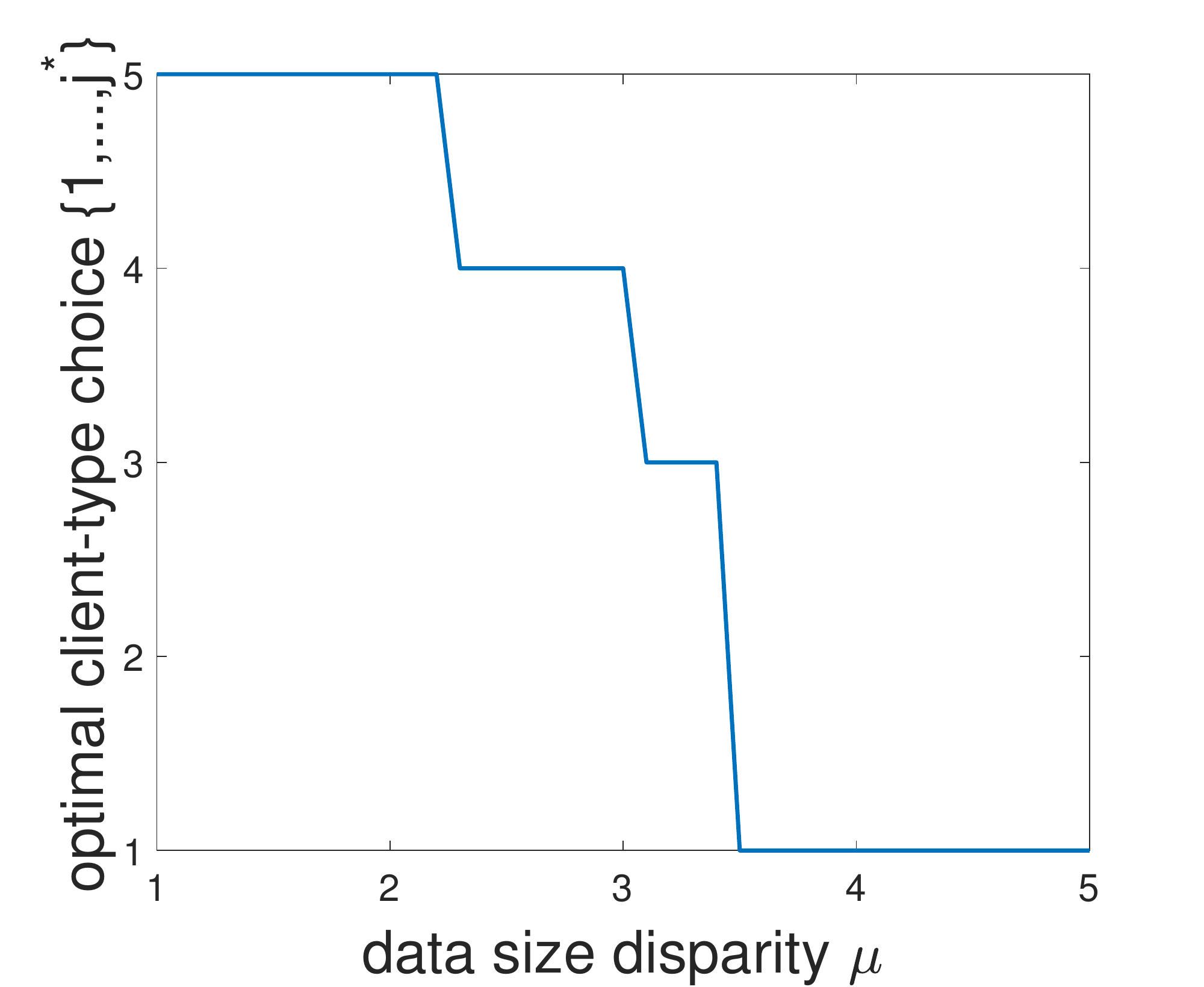}
\end{minipage}
}
\subfigure[Optimal client-type choice $j^*$ versus training rate $\frac{s}{\tau}$.]{\label{fig_Groupswithbeta}
\begin{minipage}{.37\textwidth}
\includegraphics[width=1\textwidth]{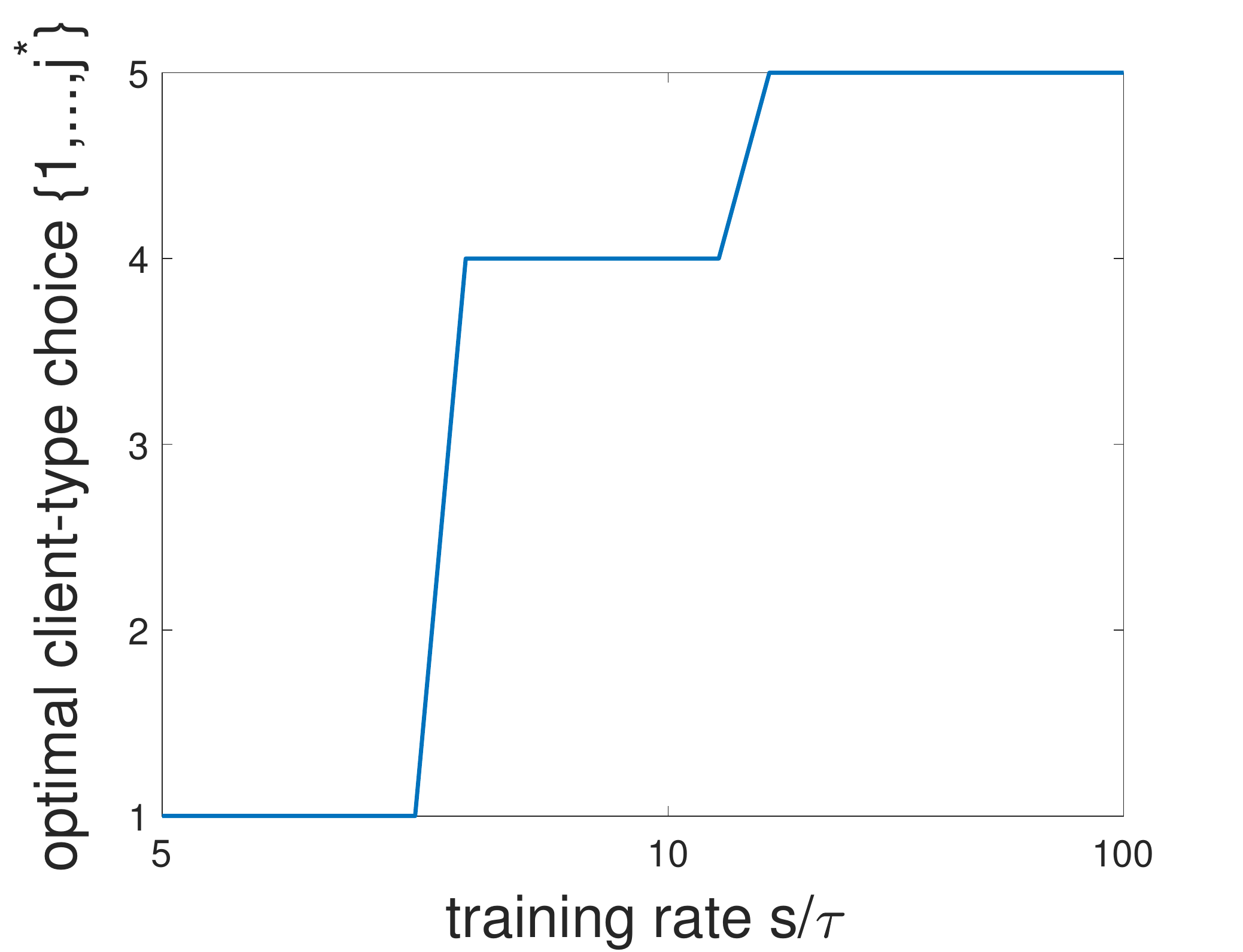}
\end{minipage}
}

\caption{Optimal client-type choice $j^*$ with $\{1,...,j^*\}$ versus data size disparity $\mu$ and training rate $\frac{s}{\tau}$.}\label{fig_Nmultiple}
\end{figure*}

\newcounter{mytempeqncnt1}
\begin{figure}
\setcounter{mytempeqncnt1}{\value{figure}}
\setcounter{figure}{4}
\centering\includegraphics[scale=0.39]{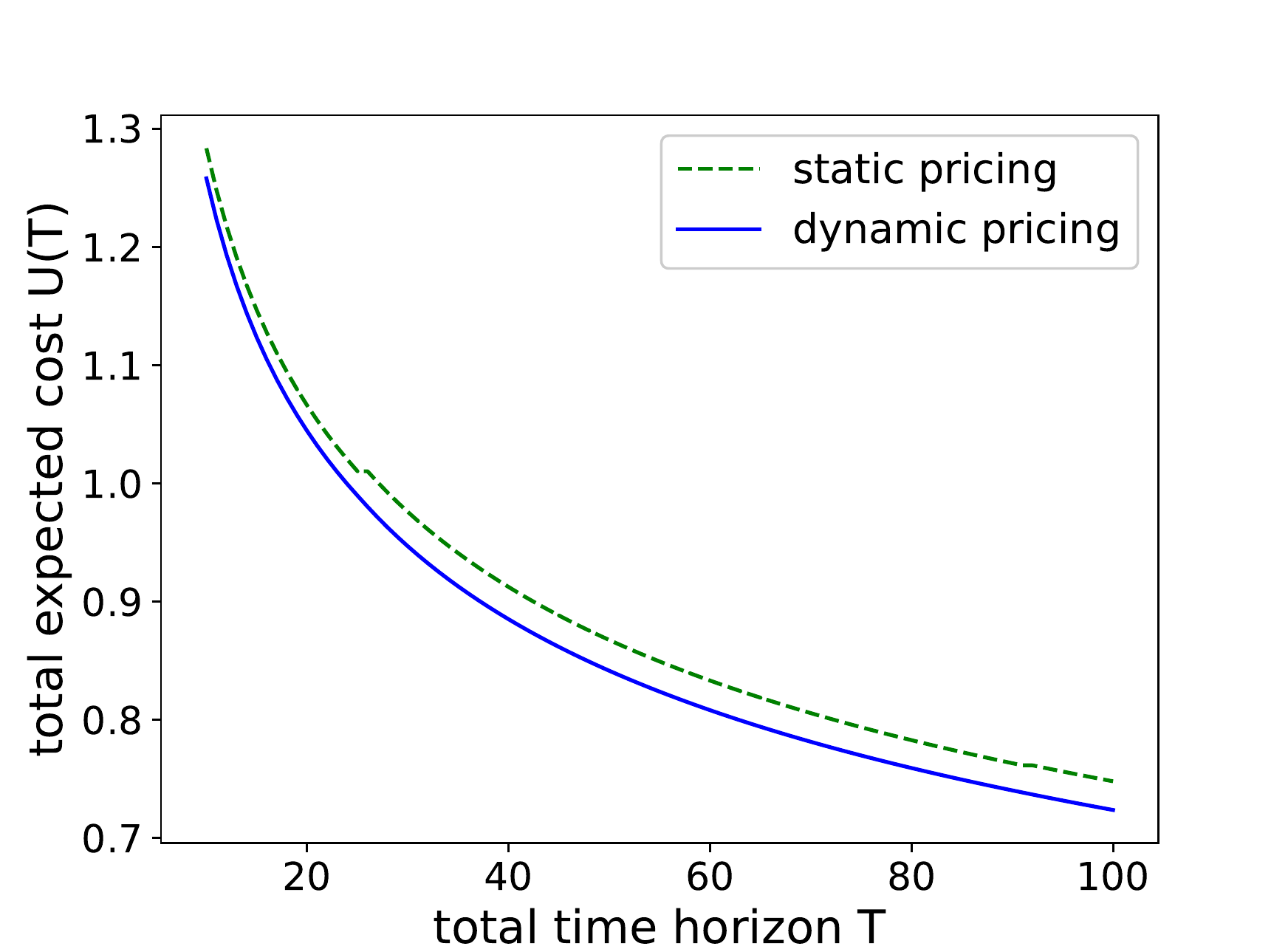}\caption{Static Pricing versus Dynamic Pricing by comparing their total expected cost objectives. }\label{fig_Static_Dynamic_Price_Loss}
\end{figure}



As shown in Proposition \ref{pro_robust}, the error term $\Phi(\delta_i|i\in\{1,...,j^{*}\})$ increases with upperbound error $\delta_i, i\in\{1,...,j^{*}\}$. Besides worst-case analysis, we also provide the average-case analysis via simulations.
As shown in Fig. \ref{fig_U_TVSDelta}, for randomly generated noisy data size $s_i(t)\in[s_i-\delta_i, s_i+\delta_i], i\in\{1,...,j^{*}\}$ in each time slot $t$, the difference between $\tilde{J}(\boldsymbol{P(t)}|T_{th}, \{1,...,j^{*}\})$ and $J(\boldsymbol{P(t)}|T_{th}$, $\{1,...,j^{*}\})$ increases with the noise bound $\delta_i, i\in\{1,...,j^{*}\}$, which coincides with the changing trend of $\Phi(\delta_i|i\in\{1,...,j^{*}\})$.


\section{Numerical experiment}\label{sec_simulations}




In this section, we conduct simulation experiments to evaluate the performance of our proposed solution. We first discuss the traditional static pricing benchmark versus our dynamic pricing for homogeneous clients with identical data size and training time. Then we consider heterogeneous clients with different data size and training time to show how different impacting factors affect optimal client-type choice.



First, we examine the performance of our proposed dynamic pricing for homogeneous clients, by comparing the total expected cost $U(T)$ under the traditional static pricing (e.g., \cite{feng2019joint, kang2019incentive}) with that under our dynamic pricing in (\ref{equ_p(t)withoutB(t)}). Note that the static pricing strategy is to use a fixed price to recruit clients at all times. The optimal static pricing can be derived as $p^{*} = min \{ (\frac{D^2b^3\tau^3(1-r)}{16T_{th}^2\alpha^3sr(1-r^{T_{th}})})^{\frac{1}{5}}$, $b(T - T_{th})\}$ according to (\ref{equ_BUniform})-(\ref{equ_U(T)}), which is a special case of (\ref{equ_p(t)withoutB(t)}). We set parameter values to be $\alpha=0.5, b=1, s=1.0, r=0.5, \tau=0.5$. The experimental results in Fig. \ref{fig_Static_Dynamic_Price_Loss} show that our proposed dynamic pricing strategy always outperforms the static pricing strategy, and the gap increases with time horizon $T$ due to the accumulated advantage of dynamic price over time. Moreover, the total expected costs $U(T)$ for both static and dynamic pricing decrease over time. This is because the model accuracy loss decreases as the number of training iterations $D$ increases with the time horizon $T$.

Then, we will show how the optimal client-type choice changes with different factors. Consider $N=5$ types of clients with uniform client distribution $\{q_i=\frac{1}{N}, i\in\{1,...,N\}\}$, different data size $\{s_i=s_0+(i-1)\mu, i\in\{1,...,N\}\}$ and training time $\{\tau_i, i\in\{1,...,N\}\}$, where $\mu$ is the data size disparity.
Note that a client's training time increases with the data size. According to \cite{tran2019federated}, we set $\tau_i=\beta s_i$, where $\beta$ is related to CPU-cycle frequency and transmission rate. Let $\alpha=0.5, b=1, s_0=1, r=0.5 \text{and } T=10$. When data size disparity $\mu=1$ and $\beta=0.01$, the optimal client-type choice is $\{1,2,3,4,5\}$ by inviting all types of clients. Starting from above setting, Fig \ref{fig_Nmultiple} shows how the optimal client-type choice changes with different impacting factors.
As the data size disparity $\mu$ between any two neighboring types increases from $1$ to $5$, it is shown in Fig. \ref{fig_GroupswithDelta} that the optimal client-type choice decreases from $\{1,2,3,4,5\}$ including all types to $\{1\}$ with only type-$1$ clients with the smallest data size and training time per iteration. As $\mu$ increases in the synchronous FL running, the clients with smaller data and training time need to wait longer for those clients with larger dataset to complete, which results in less global training iterations and thus we drop higher client types. 
As the training rate $\frac{s_i}{\tau_i}=\frac{1}{\beta}$ for each client-type $i$ increases, Fig. \ref{fig_Groupswithbeta} shows that it is better to recruit more types of clients without worrying about their training time difference. Our simulations also show that we will include more client types given a longer time horizon T and more client types $N$. We skip the details here due to the page limit.

\section{Conclusion}\label{sec_conclusion}

In this paper, we focus on the clients' incentive mechanism design in FL, by offering time-dependent monetary rewards per client arrival to trade-off between the total payment and the FL model’s accuracy loss, under incomplete information about their random arrivals and private training costs. We jointly consider two phases including the client recruitment phase and model training phase to balance the total data size and training iterations. First, for homogeneous clients with identical data size and training time, we obtain a new dynamic pricing solution in closed-form to optimally balance the total payment to clients and the accuracy loss. Such pricing scheme gradually increases when close to recruitment deadline due to aging effect. Moreover, for heterogeneous clients with different data size and training time, we use a three-stage model to successfully extend our dynamic pricing solution. A linear algorithm is proposed to find the optimal client recruitment threshold and monotonically select client types for FL. Finally, we show the robustness of our solutions against estimation error of clients' data size and run numerical experiments to validate our analytical results.


\balance
\bibliographystyle{ACM-Reference-Format}
\bibliography{Dynamic Pricing in FL.bbl}

\end{document}